\documentclass[%
 reprint,
nofootinbib,
 amsmath,amssymb,
 aps,
prd,
]{revtex4-2}

\usepackage{graphicx}% Include figure files
\usepackage{dcolumn}% Align table columns on decimal point
\usepackage{bm}% bold math
\usepackage[mathlines]{lineno}% Enable numbering of text and display math
\usepackage{xcolor}

\begin{document}

\preprint{PRELIMINARY}

%\title{Measurement of the proton and helium-4 inelastic hadronic\\ cross sections with the Dark Matter Particle Explorer}

\title{Hadronic cross section measurements with the DAMPE\\ space mission using 20~GeV--10~TeV cosmic-ray protons and $^4$He}

% Force line breaks with \\

%\author{Paul Coppin} \email{paulcppn@gmail.com}

%\affiliation{Département de physique nucléaire et corpusculaire,\\ Université de Genève, 24 quai Ernest-Ansermet,\\ Genève CH-1205, Switzerland}%
%\collaboration{DAMPE Collaboration}%\noaffiliation

%\affiliation{
%\affiliation{
%\affiliation{%


\renewcommand{\thefootnote}{\fnsymbol{footnote}}

% Redefine \fnsymbol to include more symbols
\makeatletter
\def\@fnsymbol#1{\ensuremath{\ifcase#1\or *\or \dagger\or \ddagger\or
		\mathsection\or \mathparagraph\or \|\or **\or \dagger\dagger\or
		\ddagger\ddagger\or \mathsection\mathsection\or \mathparagraph\mathparagraph\or \#\or \bullet\or \cdot\or \star\or
		\dagger*\or \ddagger*\or \mathsection*\or \mathparagraph*\or \|\or \#\or \dagger\mathsection\or
		\dagger\mathparagraph\or \dagger\|\or \dagger\#\or \dagger\bullet\or
		\dagger\cdot\or \dagger\star\or \ddagger\mathsection\or
		\ddagger\mathparagraph\or \ddagger\|\or \ddagger\#\or \ddagger\bullet\or
		\ddagger\cdot\or \ddagger\star\else\@ctrerr\fi}}
\makeatother

% tiny scriptsize footnotesize small normalsize

\author{\footnotesize F.~Alemanno$^{1,2}$,
	Q.~An$^{3,4}$,
	P.~Azzarello$^{5}$,
	F.~C.~T.~Barbato$^{6,7}$,
	P.~Bernardini$^{1,2}$,
	X.~J.~Bi$^{8,9}$,
	I.~Cagnoli$^{6,7}$,
	M.~S.~Cai$^{10,11}$,
	E.~Casilli$^{1,2}$,
	E.~Catanzani$^{12}$
% ...
	\\
	(DAMPE Collaboration)*\footnotemark[1] 
	\\
	\noindent
	$^{1}$Dipartimento di Matematica e Fisica E. De Giorgi, Università del Salento, I-73100, Lecce, Italy \\
	$^{2}$Istituto Nazionale di Fisica Nucleare (INFN) - Sezione di Lecce, I-73100, Lecce, Italy \\
	$^{3}$State Key Laboratory of Particle Detection and Electronics, University of Science and Technology of China, Hefei 230026, China \\
	$^{4}$Department of Modern Physics, University of Science and Technology of China, Hefei 230026, China \\
	$^{5}$Department of Nuclear and Particle Physics, University of Geneva, CH-1211 Geneva, Switzerland \\
	$^{6}$Gran Sasso Science Institute (GSSI), Via Iacobucci 2, I-67100 L’Aquila, Italy \\
	$^{7}$Istituto Nazionale di Fisica Nucleare (INFN) - Laboratori Nazionali del Gran Sasso, I-67100 Assergi, L’Aquila, Italy \\
	$^{8}$Institute of High Energy Physics, Chinese Academy of Sciences, Yuquan Road 19B, Beijing 100049, China \\
	$^{9}$University of Chinese Academy of Sciences, Yuquan Road 19A, Beijing 100049, China \\
	$^{10}$Key Laboratory of Dark Matter and Space Astronomy, Purple Mountain Observatory, Chinese Academy of Sciences, Nanjing 210023, China \\
	$^{11}$School of Astronomy and Space Science, University of Science and Technology of China, Hefei 230026, China \\
	$^{12}$Istituto Nazionale di Fisica Nucleare (INFN) - Sezione di Perugia, I-06123 Perugia, Italy \\
	$^{13}$Institute of Modern Physics, Chinese Academy of Sciences, Nanchang Road 509, Lanzhou 730000, China \\
	$^{14}$National Space Science Center, Chinese Academy of Sciences, Nanertiao 1, Zhongguancun, Haidian district, Beijing 100190, China \\
	$^{15}$Istituto Nazionale di Fisica Nucleare (INFN) - Sezione di Bari, I-70125, Bari, Italy \\
	$^{16}$Dipartimento di Fisica “M. Merlin” dell’Università e del Politecnico di Bari, I-70126, Bari, Italy \\
	$^{17}$Department of Physics and Laboratory for Space Research, the University of Hong Kong, Pok Fu Lam, Hong Kong SAR, China \\
}

\footnotetext[1]{Contact author: dampe@pmo.ac.cn}
\footnotetext[2]{Now at Istituto Nazionale Fisica Nucleare (INFN), Sezione di Napoli, IT-80126 Napoli, Italy.}
\footnotetext[3]{Now at Institute of Physics, Ecole Polytechnique Fédérale de Lausanne (EPFL), CH-1015 Lausanne, Switzerland.}
\footnotetext[4]{Now at Dipartimento di Fisica e Chimica “E. Segrè”, Università degli Studi di Palermo, via delle Scienze ed. 17, I-90128 Palermo, Italy.}
\footnotetext[5]{Now at Inter-university Institute for High Energies, Université Libre de Bruxelles, B-1050 Brussels, Belgium.}
\footnotetext[6]{Now at Istituto Nazionale Fisica Nucleare (INFN), I-35131 Padova, Italy.}
\footnotetext[7]{Now at New York University Abu Dhabi, United Arab Emirates.}
\footnotetext[8]{Now at Shandong Institute of Advanced Technology (SDIAT), Jinan, Shandong, 250100, China.}
\footnotetext[9]{Also at School of Physics and Electronic Engineering, Linyi University, Linyi 276000, China.}

\renewcommand*{\thefootnote}{\arabic{footnote}}

\date{\today}% It is always \today, today,

\begin{abstract}
Precise direct cosmic-ray (CR) measurements provide an important probe to study the energetic particle sources in our Galaxy, and the interstellar environment through which these particles propagate. Uncertainties on hadronic models, ion-nucleon cross sections in particular, are currently the limiting factor towards obtaining more accurate CR ion flux measurements with calorimetric space-based experiments. We present an energy-dependent measurement of the inelastic cross section of protons and helium-4 nuclei (alpha particles) on a Bi$_4$Ge$_3$O$_{12}$ target, using 88~months of data collected by the DAMPE space mission. The kinetic energy range per nucleon of the measurement points ranges from 18~GeV to 9~TeV for protons, and from 5~GeV/n to 3~TeV/n for helium-4 nuclei. Our results lead to a significant improvement of the CR flux normalisation. In the case of helium-4, these results correspond to the first cross section measurements on a heavy target material at energies above 10~GeV/n.

%\begin{description}
%\end{description}
\end{abstract}

\maketitle

%\tableofcontents

\section{\label{sec:level1}Introduction} % or JHEP, PRD, Nucl. Phys. A

Recent direct observations of Cosmic-Ray (CR) nuclei in a broad energy range, from a few GeV to hundreds of TeV, have elicited the need for a precise knowledge of hadronic cross sections.
Numerous cross section studies have been performed by accelerator experiments (see e.g. \cite{IHEP76,FNAL79,Avakian86,Jaros78,Ableev85,Tanihata85,CMSProtonProton,CMSProtonLead,LHCb13TeVProtonProton,Atlas7TeVProtonProton}). However, for ions heavier than proton, results are generally scarce and constrained to sub-GeV energies. Space-borne CR experiments present an excellent avenue to complement the low-energy measurements from accelerators. They are continually bombarded by GeV to PeV ions from proton to beyond iron. As such they can probe the cross section of various ions over a wide energy range.

The feasibility of such studies has already been demonstrated by the AMS-02 collaboration, who measured the cross sections of ions with charge $2\leq Z\leq 16$ on a carbon target \cite{AMSXS}. There is nonetheless still a strong need for analyses of complementary targets and energy ranges. Particularly heavy $\left(A\gtrsim 50\right)$ target materials, commonly present in calorimetric detectors, remain to be probed. Unlike for light targets, the conversion of cross sections from one heavy target to the next can be made in a reliable manner using e.g. the Glauber approach \cite{Glauber,Glauber0} accounting for the Gribov inelastic screen corrections \cite{Gribov1,Gribov2}. Hence, measuring the cross section for a single heavy target is sufficient to provide more accurate cross sections for the set of all heavy targets.

In this publication, we present a cross section measurement for protons and helium-4 ions (alpha particles) on a $\text{Bi}_4\text{Ge}_3\text{O}_{12}$ target using data from the Dark Matter Particle Explorer (DAMPE) satellite. These measurements present an important first step in determining the cross section for heavy targets. Additionally, they significantly contribute towards reducing the uncertainties from hadronic models; which represent one of the primary sources of systematic uncertainty in CR ion fluxes. To illustrate this point, it will be demonstrated how our results can aid to determine if discrepancies in the helium flux previously published by AMS-02~\cite{AMS7years}, ATIC-II~\cite{ATICHelium}, CALET~\cite{CALETHelium}, CREAM~\cite{CreamHelium}, and DAMPE~\cite{DAMPEHelium,ArshiaICRC} could be due to the assumption of different hadronic cross section models.

In the following, a short description will first be given of the DAMPE experiment in Section~\ref{sec:DAMPE}. Then the Geant4 and FLUKA models used to simulate CR interactions in the DAMPE detector will be described in Section~\ref{sec:Sim}, followed by an overview of the analysis methods in Section~\ref{sec:Ana}. Finally, the measurements of the inelastic hadronic cross section of proton and helium-4 on $\text{Bi}_4\text{Ge}_3\text{O}_{12}$ are presented in Section~\ref{sec:Res}, alongside a discussion on the impact of these results on the precision of CR flux measurements.

\vfill\null

\section{The Dark Matter\\ Particle Explorer}\label{sec:DAMPE}

The Dark Matter Particle Explorer (DAMPE) was launched on December 17th, 2015. The spacecraft follows a Sun-synchronous orbit around Earth at an altitude of 500~km. Its payload, which is always pointed towards the direction of zenith, consists of four subdetectors stacked in a layered design. Starting from the direction of zenith, these are: the Plastic Scintillator Detector (PSD), Silicon-Tungsten tracKer converter (STK), Bismuth-Germanium-Oxide calorimeter (BGO), and NeUtron Detector (NUD). A short overview is given below on the design and functioning of each of these four subdetectors. For a more detailed account of the detector and its in-flight operation, we refer the reader to~\cite{DampeMission,DAMPECalibration}.

\textbf{PSD.} First in line is the plastic scintillator detector, which measures the ionisation energy of incoming charged particles before they interact inelastically in the detector \cite{PSD,PSD2,Calibration1,BeamTest4}. The energy loss in PSD can be used to calculate the particle charge based on the Bethe–Bloch formula, $Z\propto \sqrt{dE/dx}$. Geometrically, the PSD consists of 82 scintillating bars. Each bar has a dimension of $10\times 28\times 884\text{ mm}^3$, and a photomultiplier tube (PMT) on either side for read-out. PSD bars are oriented in two staggered double layers, orthogonal to each other and to the direction of zenith. This design aims to maximize the detection efficiency $(>99.75\%)$, allowing PSD to serve as an anti-coincidence shield for gamma-ray observation, while at the same time providing a granularity necessary to distinguish primary CRs from backscattered particles.

\textbf{STK.} Placed after the PSD are 768 single-sided silicon micro-strip detectors (SSD) \cite{STKCalibration,Calibration2,STKCalibration2,STKCalibration3}. These SSDs are equally distributed over 12 layers, alternately along the $x$- and $y$-direction. Each SSD contains 768 strips with a width of 48~$\mu$m, length of 93.196~mm, and pitch of 121~$\mu$m \cite{STKCalibration2}. The combined signal of the SSDs enables reconstructing the trajectory of charged particles which pass through the STK. Additionally, the STK provides a charge measurement complementary to the PSD, and serves to convert photons into electron-positron pairs thanks to three 1~mm thick sheets of tungsten located between its central layers.

\textbf{BGO.} Weighting over one~ton, DAMPE houses the heaviest calorimeter of all current CR space missions \cite{BGOCalibration,BeamTest3,BGOCalibration2}. Similar to PSD and STK, BGO has a layered design with 308~bars spread over 14~layers, which alternate between the $x$- and $y$-direction. Each bar is made of a $\text{Bi}_4\text{Ge}_3\text{O}_{12}$ composite, has a dimension of $25\times 25\times 600\text{ mm}^3$, and is read out by a PMT on either side. Thanks to this extended calorimeter of 32~radiation lengths, DAMPE can probe CR ion fluxes up to kinetic energies of hundreds of TeV. Additionally, as a one~ton mass of uniform composition, BGO serves as an ideal target for the ion-nucleon cross section measurements performed in this work.
%\vfill\null

\textbf{NUD.} The last subdetector encountered by particle showers is the neutron detector \cite{Calibration4}. NUD consists of four boron-loaded plastic scintillators read out by PMTs. These scintillators enable measuring delayed neutrons from particle showers in the calorimeter, through the interaction $^{10}B + n \rightarrow {}^7{Li} + \alpha + \gamma$. Hadronic showers generally produce around $\sim$10 times more neutrons than electromagnetic showers. The main design goal of NUD is to improve proton discrimination in analyses of the electron flux. No data from NUD is used in the present analysis.

\begin{figure}[t]
	\includegraphics[width=8.6cm]{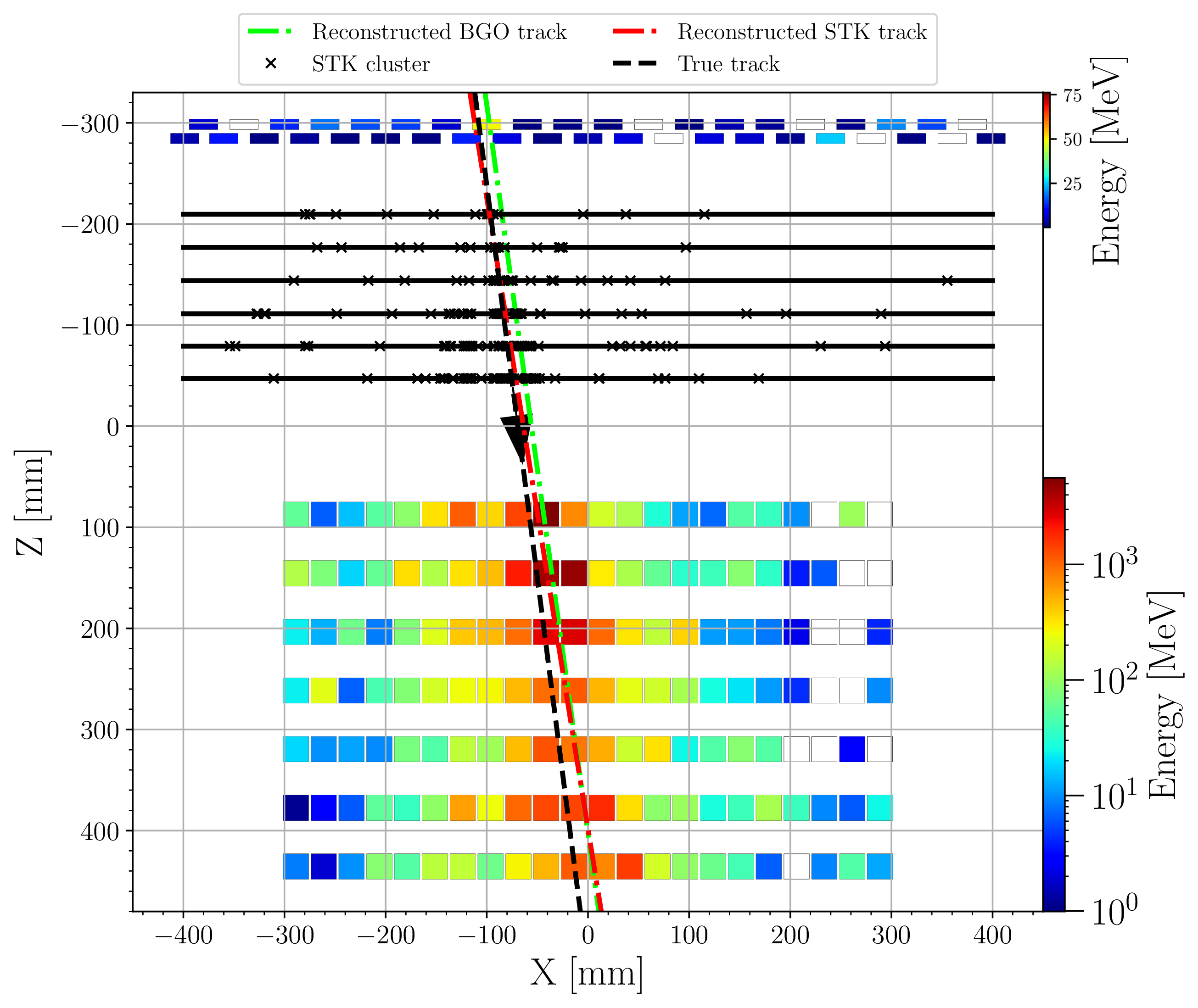}
		\caption{\label{fig:EventDisplay} Visualisation of a proton CR interacting with the DAMPE instrument, as simulated with Geant4. The three subdetectors which are used in the work: PSD, STK, and BGO; are shown from top to bottom, along with the true and reconstructed track directions.}
\end{figure}

Figure~\ref{fig:EventDisplay} shows the simulated response of the three subdetectors used in this work, projected in the $xz$-plane. Only half of all bar and strips detectors in DAMPE are depicted in the figure for clarity, namely those which are orthogonal to the projection plane.

\section{Simulation}\label{sec:Sim}
The cross section measurements presented in this work rely on the procedure of forward-folding. Specifically, data collected by the DAMPE satellite is compared to simulated event samples. By varying the inelastic hadronic cross section of the simulated events, the measured cross section is determined as that for which the observables of simulated events best match the data.

This approach necessitates an accurate modelling of the interaction of CR ions in the DAMPE experiment. Two simulation frameworks are considered for this purpose: Geant4 and FLUKA.

\textbf{Geant4.} Simulations in Geant4 are performed with version 10.05.p01 using the FTFP-BERT physics model below 100~TeV \cite{Geant41,Geant42,Geant43}. Low energy ions are simulated with the Bertini intra-nuclear cascade model (BERT). High-energy ions are treated using a combination of the Fritiof string (FTF) and Precompound/de-excitation (Preco) models. The transition between these two models occurs in the interval from 3~GeV to 12~GeV. For simulations above 100~TeV, EPOS LHC \cite{EPOS,EPOS-LHC} is used, linked to Geant4 through the CRMC framework \cite{CRMC,CRMC_Tykhonov}.

\textbf{FLUKA.} Simulations in FLUktuierende KAskade (FLUKA) are performed with version 2011.2x7 \cite{FLUKA1,FLUKA2}. Like Geant4, the hadronic model used in FLUKA depends on the considered energy range. Low-energy interactions are treated by the PreEquilibrium Approach to NUclear Thermalization (PEANUT) and the resonance production and decay model. Hadron-hadron and hadron-nucleon interactions from 5~GeV to 20~TeV rely on the Dual Parton Model (DPM) and Glauber-Gribov model with Generalised Intranuclear Cascade (GINC). For heavy ion interactions above 5~GeV/n, and hadron-hadron and hadron-nucleus interactions above 20~TeV, FLUKA interfaces to DPMJET-III \cite{dpmjet3_1,dpmjet3_2,dpmjet3_3}.

\textbf{Setup.} Simulations in both Geant4 and FLUKA model the DAMPE experiment with a detailed GDML geometry that is based on CAD drawings of the satellite \cite{STKCalibration3}. Proton and helium-4 particles are generated to represent an isotropic flux of CRs \footnote{Only down-going primary particles are simulated, as DAMPE is shielded from below by Earth.}, with kinetic energies from 10~GeV to 1~PeV. Weights are given to events such that their energy spectrum matches that of the proton and helium-4 flux measured by DAMPE \cite{DAMPEHelium,DAMPEProton,ArshiaICRC}. Following the simulation with Geant4 or FLUKA, a digitisation process is applied to simulate the electronic detector response. After this, the same reconstruction and trigger algorithms which are used for data are applied to the simulated events.

\textbf{Beam-tests.} To probe the simulation accuracy, extended beam-tests were performed at the CERN Proton Synchrotron (PS) and Super Proton Synchrotron (SPS) \cite{BeamTest1,BeamTest2,BeamTest3,BeamTest4}. These studies have shown that in terms of the BGO energy response, particularly the trigger efficiency \cite{BeamTest2}, transverse shower development \cite{BeamTest2}, and quenching effect \cite{BeamTest1}; Geant4 is generally in better agreement with beam-test data than FLUKA. Geant4 is therefore taken to be the baseline simulation framework of this analysis. Simulations using FLUKA will be used to verify the results, and test the systematic uncertainty linked to the hadronic model.

\section{Analysis}\label{sec:Ana}
%\subsection{Data sample}
Results presented in this work are based on 88~months of DAMPE data, collected from May 2016 to September 2023. Standard calibration corrections \cite{Calibration1,Calibration2,Calibration3,Calibration4} are applied to the data prior to the analysis. These corrections ensure that the detector response is uniform throughout time, independent of temperature fluctuations during the orbit and detector ageing effects.

\subsection{Event selection\label{sub:EventSelection}}
\textbf{Trigger.} The primary DAMPE trigger, used in the analysis of CR-fluxes, requires an energy deposition corresponding to 10 minimum ionising particles (MIP) in the first four layers of BGO \cite{DAMPETrigger}. This criterion aims to select particles which start showering either before or in the first BGO layer. Cross sections are measured in this work by examining the depth at which particles interact inelastically in the calorimeter. It follows that the primary trigger cannot be used for the current analysis. Instead, the so-called MIP-trigger is used, which requires an energy deposition of at least 0.4~MIPs in BGO layers 3, 11 and 13; or in BGO layers 4, 12, and 14 \cite{DAMPETrigger}. To avoid surpassing the detector downlink to Earth, the MIP-trigger is pre-scaled by a factor four and only activated when the satellite's latitude is within $\left[-20^\circ,~20^\circ\right]$, where the flux of low-energy particles is reduced due to shielding by Earth's magnetic field. Within that declination band, such shielding is significantly decreased in the South-Atlantic Anomaly (SAA). Data taken in the SAA are therefore excluded from the analysis.

\textbf{Containment.} Following the trigger, the first selection criterion applied to data is that the reconstructed track of the primary particle should traverse the PSD, STK, and BGO subdetectors from top to bottom. Side-entering or -exiting events, which only deposit a partial or no signal in any of the subdetectors, are thus removed. For this purpose, machine-learning (ML) based tracking is applied using data from the BGO and STK subdetectors \cite{AndriiTracking}. Cases in which producing a reliable track estimate is most challenging occur mainly for horizontal events, and tracks which graze the edge of the detector. If the tracking fails to produce an accurate result, such events could erroneously be accepted as down-going contained events. To remove such sources of background, additional cuts are applied. Specifically, events are rejected if:
\begin{itemize}\addtolength\itemsep{-2.5mm}
	\item more than 35\% of the total BGO energy is contained within a single layer,
	\item the bar which has the maximal energy in BGO layer 1, 2, or 3 is one of the outermost bars,
	\item a one-dimensional $\chi^2$ fit of the deposited BGO energy in the $x$- and $y$-direction fails, or when it does not intersect the top and bottom of the calorimeter,
	\item No energy was deposited in any of the PSD bars intersecting the reconstructed primary track.
\end{itemize}

\textbf{Lepton rejection.} At low energies, leptons form a small ($\sim$3\%) but non-negligible background for proton analyses. The dominant background flux comes from CR electrons. Since DAMPE cannot measure the sign of charged particles, electrons leave a signal almost identical to that of proton in the PSD and STK subdetectors. However, their differentiation can be made using the calorimeter. Leptons induce electromagnetic showers are on average more dense than hadronic showers. A variable, $\zeta$, that probes the shower collimation and penetration power has been developed for analyses of the CR electron flux~\cite{DampeElectronNature}. In this analysis, a cut is imposed on $\zeta$ which reduces the electron background by $>99\%$ while retaining $>99\%$ of ion events, as detailed in Appendix~\ref{app:LeptonRejection}.

\textbf{Ion selection.} For proton and helium, the charge identification can be performed using either the signal from PSD or STK. Silicon strips in STK have an effective spatial readout dimension which is more than two orders of magnitude smaller than that of PSD. As a result, charge measurements with STK are much less affected by backscattered particles. Proton and helium are therefore selected in this work based on the signal measured by the STK subdetector. On the order of 6\% (14\%) of proton (helium) particles will interact inelastically before they reach the STK. In such case, no reliable charge measurement can be made by the STK. These events interacting early-on are rejected using an ML-based classifier (see \cite{AndriiTracking} for details).

\begin{figure}[b]
	\includegraphics[width=8.6cm]{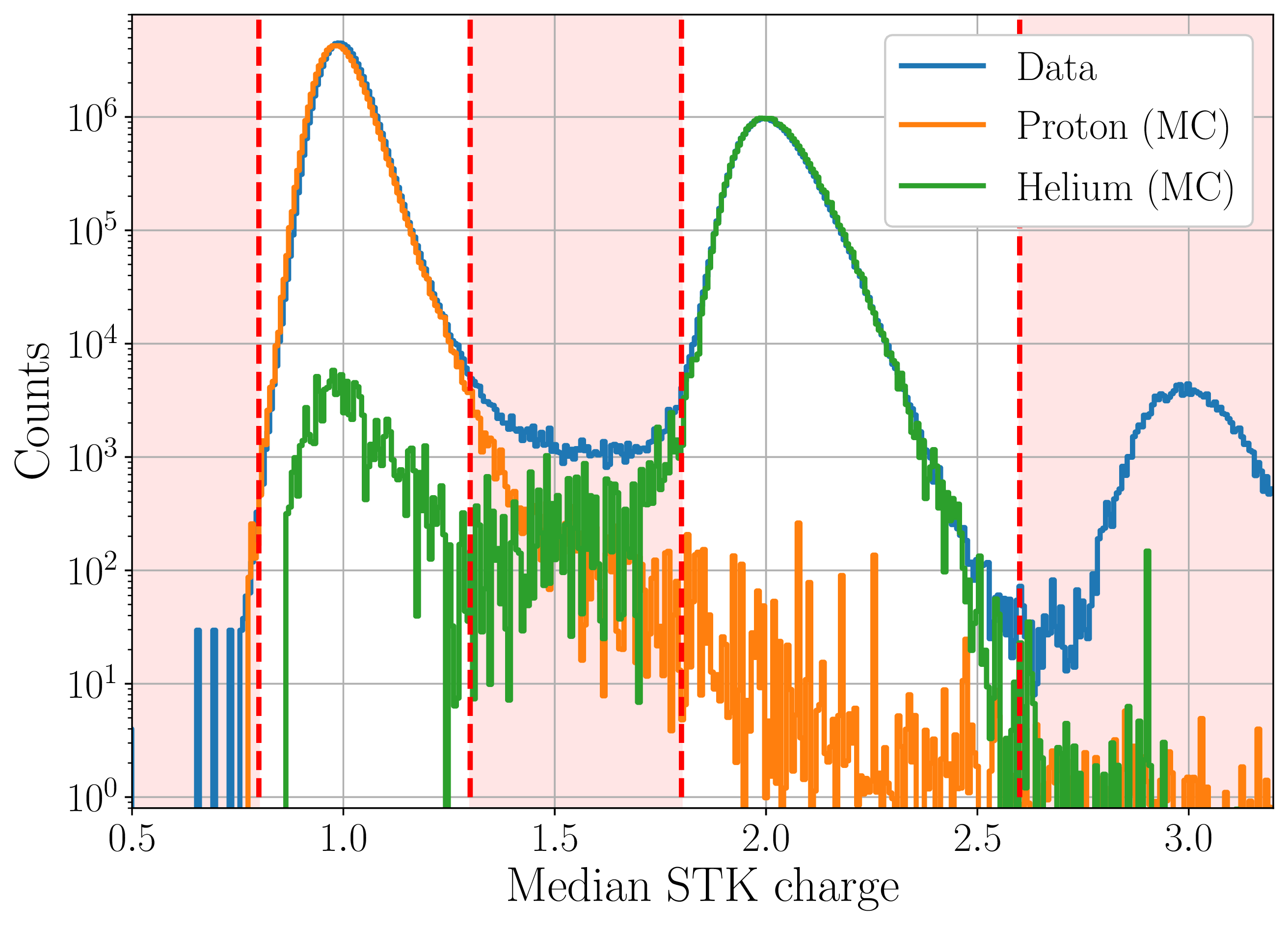}
	\caption{\label{fig:EventSelection} Distribution of the median STK charge of data and Geant4 simulation after applying the event selection cuts. The vertical red lines indicate the charge window used to select proton and helium events.}
\end{figure}

The STK provides at most 12 estimates of the primary CR charge, i.e. one for each layer in which the track passes through an SSD. The median of all non-zero charges is said to be reliable if there are at least six or more layers in which the charge deviates by less than 0.3 from the median. Figure~\ref{fig:EventSelection} show the charge distribution of events which pass this criterion. The contribution from proton and helium are clearly distinguishable, and a good agreement between data and simulation is observed. A charge window from $\left[0.8,~1.3\right]$ and $\left[1.8,~2.6\right]$ is used to select proton and helium, respectively. The combined efficiency of the STK charge cuts is $\geq85\%$ for proton and helium. After applying all selection criteria, the background from events not induced by proton or helium is estimated to be $<0.2\%$ in both cases.

\subsection{Interaction depth classifier}
To measure the inelastic hadronic cross section, the depth in the calorimeter that corresponds to the start of the hadronic shower is determined for every event. This depth can be resolved roughly up to the granularity of the detector. Events are therefore classified into 16 categories. There is one class for events which interact before BGO, one class for every layer of the calorimeter, and one class for events which pass through the entire calorimeter without interacting inelastically.

The classification is done using a boosted decision trees classifier, based on the XGBoost (XGB) library \cite{XGB}. As training input, the XGB classifier is provided with a mixture of proton and helium events simulated with both Geant4 and FLUKA. Input variables for the classifier are primarily based on the rationale that, prior to the inelastic interaction, a MIP-like signal is deposited along the primary particle's track. Additional signals can be observed above the inelastic interaction due to phenomena such as Bremsstrahlung photon production or particles from the shower backscattering upstream. However, these are usually not directly parallel to the primary's track, and limited compared to the actual hadronic shower development. Once the inelastic interaction takes place, a growing energy deposition is generally observed as a function of depth up to the shower maximum. These features led to the identification of the following 70 input variables used in the classifier:
\begin{itemize}\addtolength\itemsep{-2.5mm}
	\item 1-14. energy per BGO layer in $\log_{10}$;
	\item 15-28. energy per BGO layer in $\log_{10}$, excluding the bar(s) through which the track passes;
	\item 29-42. energy per unit distance per BGO layer, for the bar(s) through which the track passes;
	\item 43-56. shower spread in each BGO layer, defined as the RMS of the charge per Eq.~\eqref{eq:RMS};
	\item 57-68. charge in every STK layer;
	\item 69. median STK charge, to easily distinguish the primary particle's type;
	\item 70. reconstructed zenith angle of the primary, to account for zenithal dependence.
\end{itemize}

An in-depth overview on the training and performance of this classifier is provided in the Appendix~\ref{app:Vertex}. Here, we limit the discussion to stating that the model has an accuracy of 80\% or higher, for proton and helium-4 simulated with either Geant4 or FLUKA. For the sake of brevity, the point at which a CR ion is predicted to have its first inelastic interaction will be referred to as the point at which the event interacts in the following.

\subsection{Likelihood}
\textbf{Ratio.}
Using the results of the XGB classifier, a comparison can be made between the depth at which CR ions interact in data and Monte Carlo (MC) simulations. For this purpose, the following fraction is considered:
\begin{equation}
	\alpha_i = \frac{N_i}{\sum_{j=2}^{10}N_j}\ \ \ \ \ \ \ \ 2\leq i \leq 10,
	\label{eq:alpha}
\end{equation}
where $N_i$ is the number of events interacting in BGO-layer $i$ for $2\leq i\leq 9$. To pool statistics, $N_{10}$ is taken to be the sum of events interacting in layer 10 through 14, and events which pass through the calorimeter without interacting. Due to the gap in read-out detectors between STK and BGO, it is challenging to distinguish events which interact just before or inside the first BGO layer. For this reason, we consider $i\geq 2$. Finally, it is worth noting that normalizing $N_i$ by the total number of events which reach the second layer of BGO without interacting, ensures that our result is not dependent on the potential mismodeling of the cross sections in detector material upstream of the calorimeter.

\begin{figure}[t]
	\includegraphics[width=8.6cm]{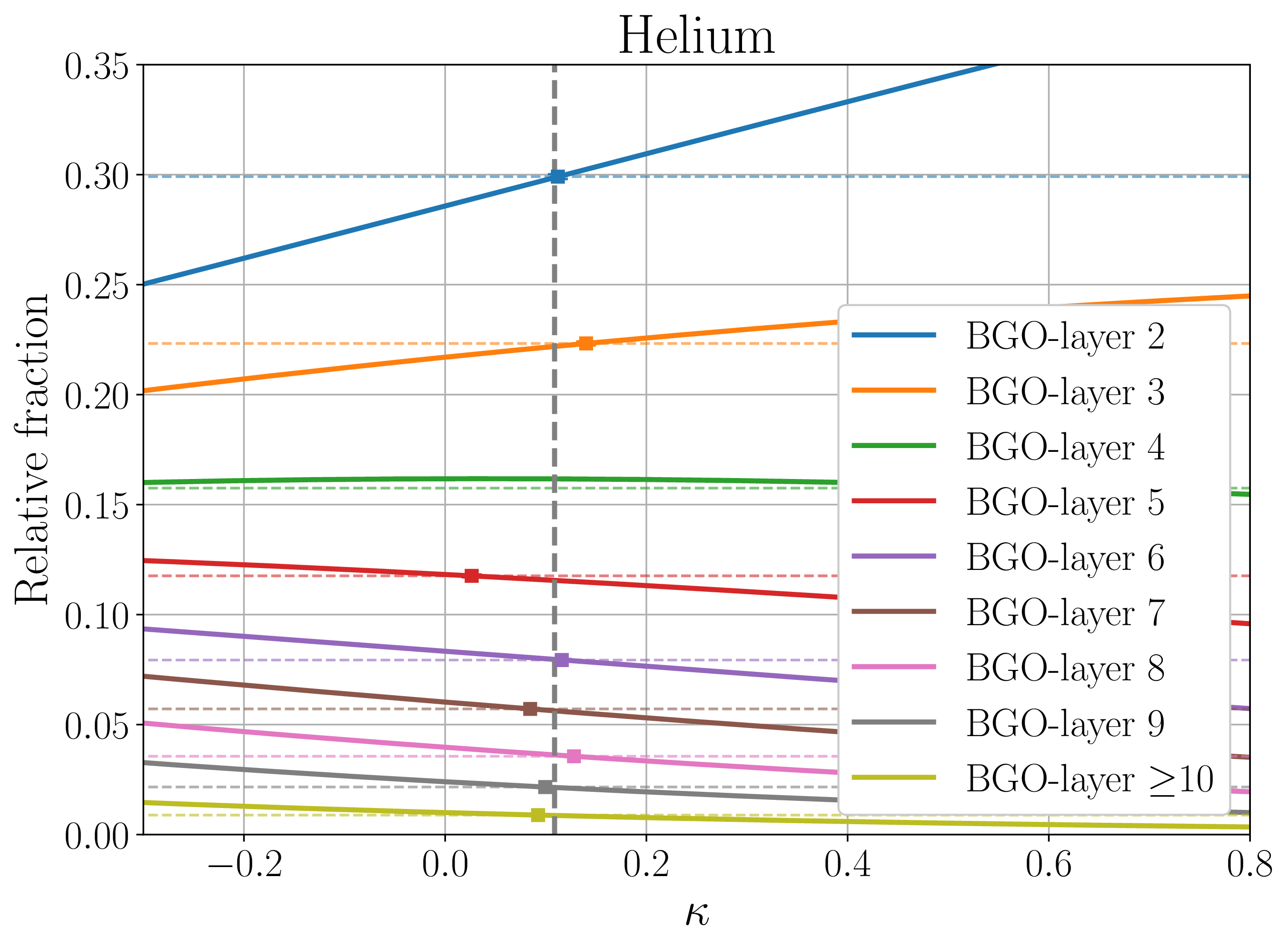}
	\caption{\label{fig:ExampleFit} Determination of the rescaling factor, $\kappa$, for helium-4 events simulated with Geant4 with deposited energies between 100~GeV and 316~GeV. Full lines show the value of $\alpha_i$, i.e. the fraction of simulated events interacting in layer-$i$ (see Eq.~\ref{eq:alpha}). Horizontal dashed lines show the true value, $N_i/N_{tot}$, extracted from data. A square is drawn at the point where data and MC intersect, indicating the best-fit value for that layer. The value obtained from the likelihood fit is indicated by the vertical dashed line.}
\end{figure}

\textbf{Probability.} 
The probability to observe $N_i$ counts in data is given by the multinomial distribution:
\begin{equation}
	\mathcal{L}=\frac{N_{tot}!}{N_2!N_3!  \cdots N_{10}!} \prod_{i=2}^{10}{ \alpha_i^{N_i} }\ ,
	\label{eq:Likelihood}
\end{equation}
where $\alpha_i$ is derived from simulation based on Eq.~\eqref{eq:alpha}, and $N_{tot}=\sum_{j=2}^{10}N_j$ is the total number of events in data that go into the likelihood. We consider as hypothesis that the true cross section is that of MC modified by a constant normalisation factor,
\begin{equation}
	\sigma_{true}(E) = (1+\kappa )\cdot \sigma_{MC}(E).
	\label{eq:Scaling}
\end{equation}
Under this assumption, the fractions $\alpha_i$ become a function of the cross section correction factor, $\kappa$. This change of $\alpha_i(\kappa)$ modifies the value of the likelihood in Eq.~\eqref{eq:Likelihood}. To determine the value of the scaling factor $\kappa$ which provides the best match to data, we start by evaluating the value of the likelihood over a grid of $\kappa$ with a resolution of 0.01. A second order polynomial fit is then performed around the maximum to determine the exact value of $\kappa$ that maximises the likelihood. A visualisation of this procedure is shown in Figure ~\ref{fig:ExampleFit}.

\subsection{Re-weighting MC}
Calculating the values of $\alpha_i(\kappa)$ and $\mathcal{L}(\kappa)$ requires simulation samples in which the cross section has been varied from the baseline value. Reproducing all existing proton and helium simulation over grid of $\kappa$ is very computationally expensive. An alternative approach is therefore used in which already existing simulation is re-weighted. Changing the cross section in MC essentially comes down to shifting the number of events that interact inelastically as a function of depth in the detector, $dN/dz$. A method has been developed to parametrise the distribution of $dN/dz$ as a function of cross section. This enables values of $\alpha_i$ to be calculated without having to repeat existing simulations. For full details on the re-weighting procedure, we refer the reader to the Appendix~\ref{app:Reweight}.

\subsection{Energy dependence}
Hadronic inelastic cross sections are measured in this work as a function of the kinetic energy per nucleon of the incident particle. The kinetic energy of recorded events is probed based on the energy deposited in the BGO calorimeter. Hadronic showers deposit on average one third of the primary kinetic energy in the calorimeter. This fraction can be significantly lower for ions with energies in excess of several tens of TeV. At those energies BGO bars can become saturated, resulting in a null read-out of affected bars. A ML-based correction \cite{MishaSaturation} is applied to those events to recover the actual deposited energy. 

\begin{figure}[t]
	\includegraphics[width=8.6cm]{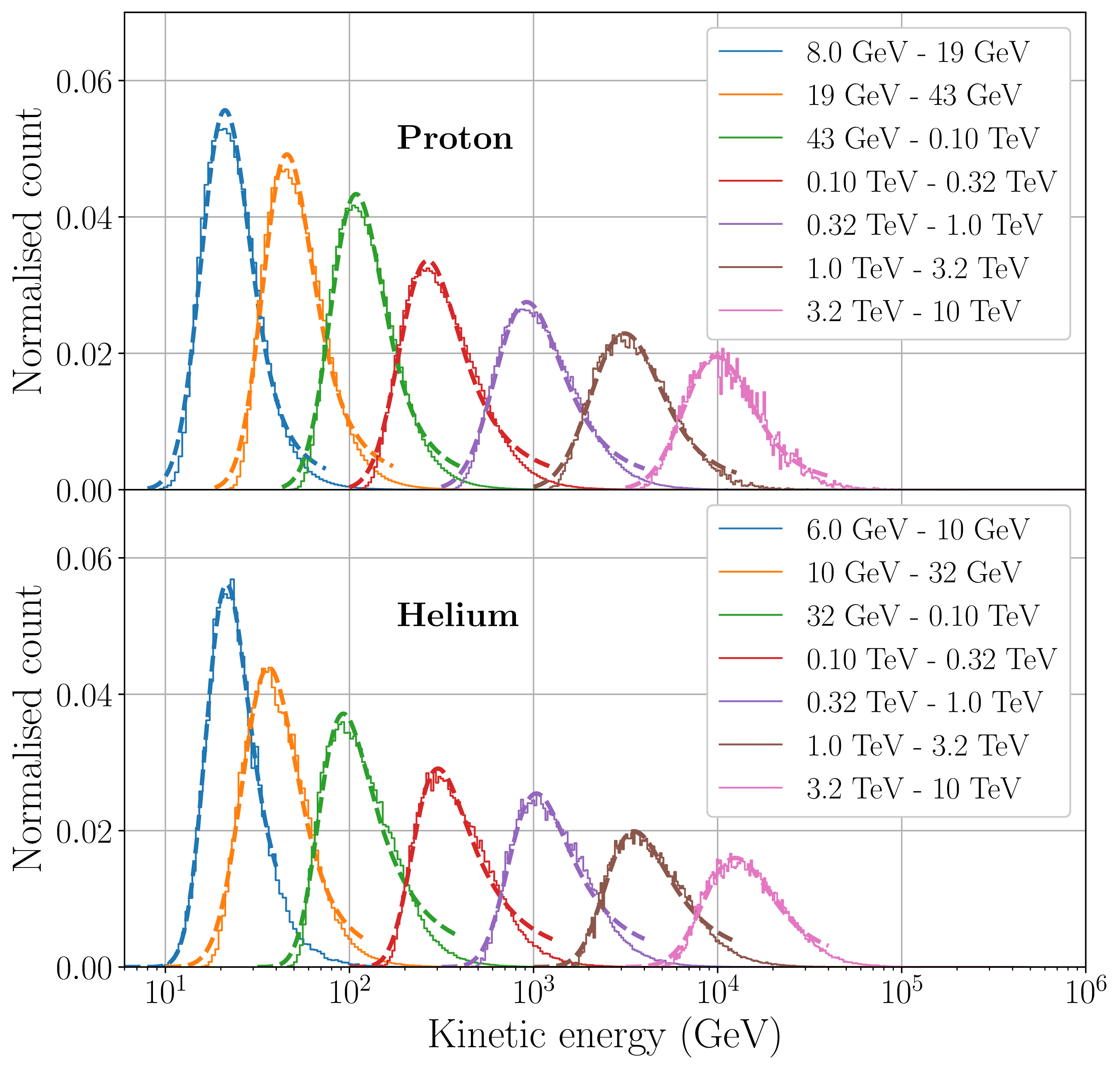}
	\caption{\label{fig:EDistribution} Full lines show the distribution in kinetic energy of proton (top) and helium-4 (bottom) events simulated with Geant4. Each colour corresponds to a different bin in deposited energy as indicated by the legend. A Landau distribution convoluted with a Gaussian is fit to every bin, shown by the dashed lines. These fits determine the reference (peak) value and uncertainty (width) for the kinetic energies corresponding to each deposited energy bin.}
\end{figure}

Proton and helium-4 candidate events are binned based on their total deposited energy corrected for saturation. A cross section measurement is performed independently on the data in every bin. Seven bins are used, ranging from 8~GeV to 10~TeV for proton, and 6~GeV to 10~TeV for helium. The  lower limit in deposited energy is chosen such that the kinetic energy of primary particles satisfies $E_{kin}\geq 10$~GeV. The upper limit is determined by statistics, as will be discussed in detail later on. Figure~\ref{fig:EDistribution} shows the kinetic energy distribution for each bin of proton and helium-4. Good agreement is observed when modelling the kinetic energy distributions with the convolution of a Landau and Gaussian density function. The central kinetic energy at which the cross section is measured is taken to be the most probable (peak) value of the fitted distribution. As a measure of variance, the width of the Landau and Gaussian component are added in quadrature.

\subsection{Uncertainty}\label{sub:uncertainty}
\textbf{Statistics.} Before presenting the measured results, an overview is given on the analysis uncertainties. A statistical error arises from the use of data counts $N_i$ in the likelihood. To estimate the effect on our measurement, pseudo-experiments are realized in which the observed counts $N_i$ are varied by binomial errors, while keeping the total number of events $N_{tot}$ constant. Ten thousand pseudo-trials are realised, each resulting in a cross section estimate. The statistical error is taken to be the standard deviation of the cross sections obtained from these pseudo trials.

Due to the rapidly dropping CR spectrum, roughly $dN/dE\propto E^{-2.7}$, the statistical uncertainty quickly increases as a function of energy. While below 0.1\% at the lowest energies, it accounts for a relative uncertainty of 20\% for proton and 13\% for helium at the highest-energy point. Statistics thus determine the maximal energy up to which the analysis is sensitive.

\textbf{Analysis procedure.} Besides statistical uncertainties, there are systematic uncertainties related to the assumptions used in the analysis procedure. An overview of these uncertainties is shown in Table~\ref{tab:Systematics}. Every systematic is evaluated separately in each of the seven energy bins. For the sake of conciseness, the discussion presented here will focus on the mean values of every category in the case of proton. For energy-dependent systematic uncertainties please see the Appendix~\ref{sup:Uncertainties}.

%The dominant systematic uncertainty of 3.3\% comes from the \textit{interaction depth classifier}. This uncertainty was quantified by developing a separate convolutional neural network (CNN) based classifier which uses an image of the calorimeter as input. The choice for using the XGB classifier as the baseline for the analysis is that it offers a larger prediction accuracy than the CNN. The second largest contribution relates to \textit{diffractive processes}. As they are characterised by small energy losses, the interaction depth classifier has a reduced accuracy of $\sim45\%$ for this subset of interactions. Due to the limited accuracy of the classifier and phenomenological uncertainties in the modelling of hadron-nucleus diffractive processes, an uncertainty of 3\% is chosen. This corresponds to the fraction of diffractive inelastic events estimated from Geant4 and FLUKA.

The dominant contribution relates to \textit{diffractive processes}. As they are characterised by small energy losses, the interaction depth classifier has a reduced accuracy of $\sim40\%$ for this subset of interactions. Due to the limited accuracy of the classifier and phenomenological uncertainties in the modelling of hadron-nucleus diffractive processes, an uncertainty of 3.5\% is chosen. This corresponds to the fraction of inelastic events that are diffractive based on the hadronic models of Geant4 and FLUKA. The second largest contribution of 3.3\% comes from the \textit{interaction depth classifier}. This uncertainty was quantified by developing a separate convolutional neural network (CNN) based classifier which uses an image of the calorimeter as input. The choice for using the XGB classifier as the baseline for the analysis is that it offers a larger prediction accuracy than the CNN. Third in line is the uncertainty of 2.6\% from the assumed \textit{CR spectrum} used to weight simulation. At energies below a few 100~GeV, excellent agreement between CR flux measurements of different experiments \cite{AMS7years,CALETProton,ArshiaICRC} allows for an uncertainty on the spectral index of 0.01 for proton and 0.05 for helium. At larger energies this uncertainty is increased to 0.1 for both proton and helium. The fourth largest contribution to the uncertainty relates to the \textit{event selection}. Cuts imposed on the STK charge and the parameters used for the lepton rejection and ion selection were loosened or restricted. These adjustments effectively increase or decrease the background contamination, corresponding to an uncertainty of 2.3\%. An important check was to repeat the analysis using events simulated through FLUKA (\textit{MC generator}) rather than Geant4. Changing the hadronic framework influenced the measurement by less than 1.5\%, demonstrating that the results show only limited dependence on the assumed hadronic interaction model. A final source of systematics comes from the absolute energy scale. It has been demonstrated in previous works \cite{BGOECAL,EresponseLinearity} that the DAMPE energy scale matches hadronic Geant4 simulations to within 1.5\%. Implementing such a shift influenced the resulting cross section on average less than 1\%.

\begin{table} %Star makes the table span the full page width
	\caption{\label{tab:Systematics} Overview of the systematic uncertainties (see main text for details). Uncertainties are evaluated separately for the seven energy bins. The values denoted in this table correspond to the median of each category.}
	\begin{ruledtabular}
		\begin{tabular}{lcc}
			Variable&
			\multicolumn{1}{c}{\textrm{Proton}}&
			\multicolumn{1}{c}{\textrm{Helium}}\\
			\hline
			Energy scale & 0.7\% & 0.6\% \\
			Isotopes & $<$0.9\% & $<$1.2\% \\
			MC generator & 1.5\% & 1.2\% \\
			Event selection & 2.3\% & 1.0\% \\
			CR spectrum & 2.6\% & 2.0\% \\
			Interaction depth classifier & 3.3\% & 3.2\% \\
			Diffraction & 3.5\% & --- \\
		\end{tabular}
	\end{ruledtabular}
\end{table}

\textbf{Isotopes.} The experimental observables to which DAMPE is sensitive are the absolute charge and total kinetic energy of CR particles. No distinction can thus be made between isotopes of the same kinetic energy. Cosmic-ray deuteron therefore forms an irreducible background in the proton analysis. AMS-02 has measured the flux ratio of CR deuteron over proton up to a kinetic energy of 20~GeV~\cite{AMSDeuteron}. The flux ratio as a function of the total kinetic energy is less than 2.7\% at 20~GeV, and shows a constant trend. Considering a conservative 3\% background of deuteron, with a cross section that is $\sim30$\% higher than that of proton \cite{FLUKA1,FLUKA2} would lead to an average cross section that is 0.9\% higher than that of pure proton. No correction is made to the measured result due to the unknown background fraction at the energies considered in our analysis. The contamination by deuteron is instead included as a systematic error.

A similar situation holds for the cross section measurement of helium-4, which is contaminated by helium-3 ions. An extrapolation of measurements by AMS-02 \cite{AMS-He3} shows that the helium-3 to helium-4 flux ratio as a function of kinetic energy is 12\% at 20~GeV, dropping to below 4\% above 1~TeV. Assuming a cross section which is 10\% lower \cite{FLUKA1,FLUKA2}, and a conservative background of 12\%, the effect on the cross section measurement is estimated to be 1.2\%. As for proton, this background is considered as an additional systematic error.

\section{Results and Discussion}\label{sec:Res}
Table~\ref{tab:Results} lists an overview of the analysis results. Cross sections presented in this work correspond to the inelastic cross section on a Bi$_4$Ge$_3$O$_{12}$ target. The measurement includes all inelastic processes with the exception of quasi-elastic scattering. Since the measurement is less sensitive to diffraction events, their contribution could be over- or underestimated. This is reflected in the systematic uncertainties in the case of proton, as described in Section \ref{sec:Ana}. In the case of helium-4, diffraction is assumed to be negligible, consistent with the predictions by Geant4 and FLUKA. The normalisation of the cross section is taken such that it describes the average probability for a primary particle to interact with a single nucleus of the Bi$_4$Ge$_3$O$_{12}$ target. A more detailed discussion is now presented, focussing first on proton and then helium-4.

\begin{figure}[t]
	\includegraphics[width=8.6cm]{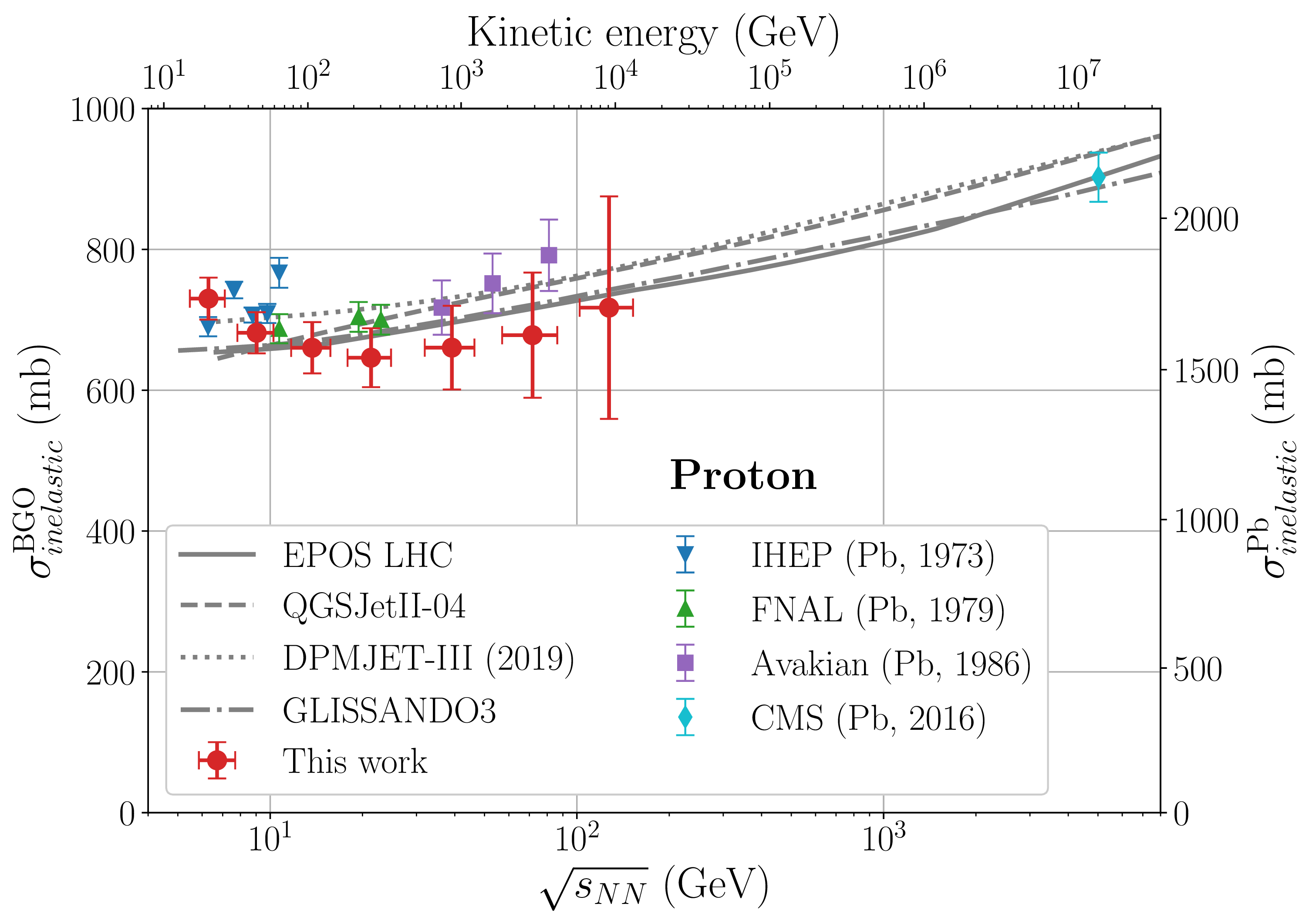}
	\caption{\label{fig:ResultProton} Inelastic cross section of proton on Bi$_4$Ge$_3$O$_{12}$ measured as a function of the center-of-mass energy per nucleon. Model predictions from EPOS LHC \cite{EPOS,EPOS-LHC}, QGSJetII-04, DPMJET-III  \cite{dpmjet3_1,dpmjet3_2,dpmjet3_3}, and GLISSANDO3 \cite{Glissando3} are shown for comparison. Additionally, previous measurements for proton on lead are shown \cite{IHEP76,FNAL79,Avakian86,CMSProtonLead}, scaled to the BGO cross section as explained in the main text. Error bars include both the statistical and the systematic error.}
\end{figure}

\textbf{Proton.} Figure~\ref{fig:ResultProton} visualises the proton cross section measurement, and provides a comparison to model predictions and previous results. No measurements have previously been published for the cross section of proton on Bi$_4$Ge$_3$O$_{12}$. Hence, we compare our results to measurements for an alternative heavy target material. Lead is chosen for this purpose, as it has a relatively well-studied cross section. To enable comparing the cross section of different materials, measurements for proton on lead are scaled by the cross section ratio between Bi$_4$Ge$_3$O$_{12}$ and lead in the EPOS LHC model \cite{EPOS,EPOS-LHC}. QGSJetII-04 \cite{QGSJET,QGSJET2}, DPMJET-III  \cite{dpmjet3_1,dpmjet3_2,dpmjet3_3} and GLISSANDO3 \cite{Glissando3} were also considered for the purpose of rescaling. They resulted in rescaling factors which are within 3\% of EPOS LHC over the considered energy range. EPOS LHC was chosen for the final comparison since its correction factor lies in between that of QGSJetII-04, DPMJET-III and GLISSANDO3. The small difference ($<3\%$) depending on the chosen models does not influence the interpretation of our results.

\begin{table} %Star makes the table span the full page width
	\caption{\label{tab:Results} Measurement of the inelastic hadronic cross section of proton and helium-4 ions on Bi$_4$Ge$_3$O$_{12}$.}
	\begin{ruledtabular}
		\begin{tabular}{cD{,}{\,\pm\,}{9}D{,}{\,\pm\,}{16}}
			Sample&
			\multicolumn{1}{c}{\textrm{$E_{kin}/n$ (GeV)}}&
			%\multicolumn{1}{c}{\textrm{$\sigma_I\pm [\text{stat}] \pm [\text{sys}]$ (mb)}}\\
			%\sigma_I, [\text{stat}], [\text{sys}] (mb)\\
			\parbox[l]{2em}{$\sigma_I$}, \parbox[l]{2.53em}{[\text{stat}]}\pm \parbox[l]{2.5em}{[\text{sys}]} (mb)\\
			\hline
            Proton  & ( 19.2, 5.9 ) \times 10^0 & 730,\parbox[l]{2.5em}{3}\pm \parbox[l]{2em}{30}\\ 
            & ( 4.2, 1.3 ) \times 10^1 & 682,\parbox[l]{2.5em}{3}\pm \parbox[l]{2em}{29}\\ 
            & ( 9.8, 3.1 ) \times 10^1 & 660,\parbox[l]{2.5em}{6}\pm \parbox[l]{2em}{36}\\ 
            & ( 24.0, 8.5 ) \times 10^1 & 646,\parbox[l]{2.5em}{7}\pm \parbox[l]{2em}{41}\\ 
            & ( 8.1, 3.3 ) \times 10^2 & 661,\parbox[l]{2.5em}{19}\pm \parbox[l]{2em}{56}\\ 
            & ( 2.7, 1.2 ) \times 10^3 & 678,\parbox[l]{2.5em}{52}\pm \parbox[l]{2em}{72}\\ 
            & ( 8.6, 3.8 ) \times 10^3 & 717,\parbox[l]{2.5em}{143}\pm \parbox[l]{2em}{67}\\ \hline
            Helium-4  & ( 5.0, 1.2 ) \times 10^0 & \parbox[l]{2em}{1077},\parbox[l]{2.35em}{6}\pm \parbox[l]{2em}{25}\\ 
			& ( 8.2, 2.8 ) \times 10^0 & \parbox[l]{2em}{1105},\parbox[l]{2.35em}{4}\pm \parbox[l]{2em}{34}\\ 
			& ( 21.8, 6.3 ) \times 10^0 & \parbox[l]{2em}{1102},\parbox[l]{2.35em}{6}\pm \parbox[l]{2em}{41}\\ 
			& ( 7.1, 2.1 ) \times 10^1 & \parbox[l]{2em}{1137},\parbox[l]{2.35em}{10}\pm \parbox[l]{2em}{50}\\ 
			& ( 24.0, 8.0 ) \times 10^1 & \parbox[l]{2em}{1166},\parbox[l]{2.35em}{28}\pm \parbox[l]{2em}{62}\\ 
			& ( 8.1, 2.8 ) \times 10^2 & \parbox[l]{2em}{1240},\parbox[l]{2.35em}{70}\pm \parbox[l]{2em}{79}\\ 
			& ( 2.8, 1.1 ) \times 10^3 & \parbox[l]{2em}{1318},\parbox[l]{2.35em}{169}\pm \parbox[l]{2em}{86}\\
		\end{tabular}
	\end{ruledtabular}
\end{table}

Our analysis points generally overlap with previous measurements for lead \cite{IHEP76,FNAL79,Avakian86,CMSProtonLead} within the uncertainty range, yet are observed to be systematically lower. It follows that the model predictions tend to overshoot our results as they are tuned to previous measurements. A curious observation is that a recent (2016) measurement by the CMS collaboration \cite{CMSProtonLead} is also lower than the value expected based on prior accelerator measurements and the expected energy dependence of cross section. Our results are thus more in line with the CMS measurement. At energies $\sqrt{s_{NN}}$ below 10~GeV, the three models shown in Fig.~\ref{fig:ResultProton} reach the limit of their validity. This is indicated by the observed deviation in energy evolution between the model predictions and measurements at these energies, and thus is no real source of discrepancy. Comparing directly to model predictions, EPOS LHC and GLISSANDO3 are observed to show the best agreement due to their slightly lower normalisation compared QGSJetII-04 and DPMJET-III.

\begin{figure}[t]
	\includegraphics[width=8.6cm]{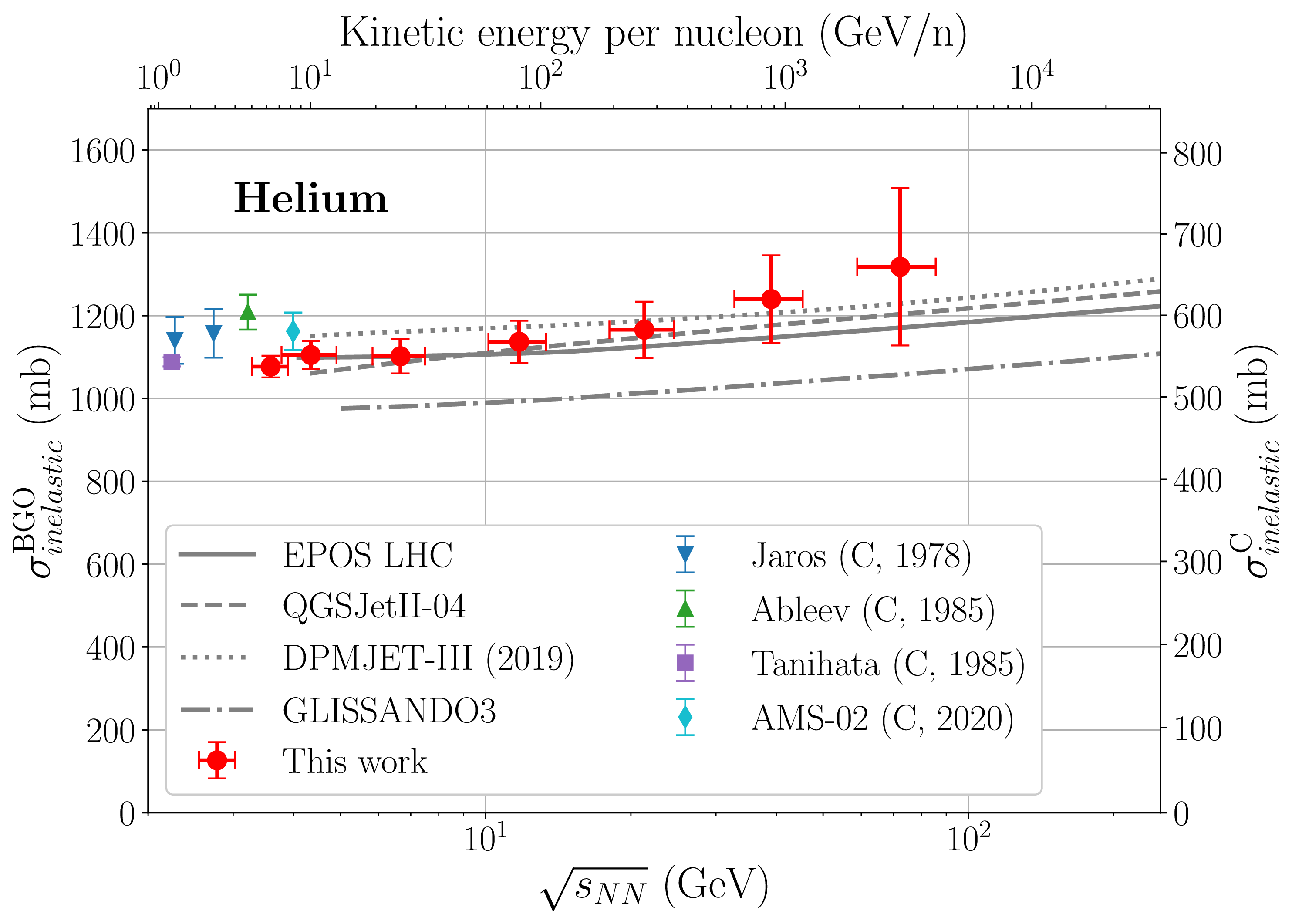}
	\caption{\label{fig:ResultHelium} Results of the energy dependent analysis for helium-4 ions on Bi$_4$Ge$_3$O$_{12}$, compared to previous measurements \cite{Jaros78,Ableev85,Tanihata85,AMSXS} for carbon targets (scaled) and model predictions \cite{EPOS,EPOS-LHC,dpmjet3_1,dpmjet3_2,dpmjet3_3,Glissando3}. Error bars include the statistical and systematic uncertainty.}
\end{figure}

\textbf{Helium.} In the case of helium-4, our results provide the first cross section measurement on a heavy target material at kinetic energies above 10~GeV/n. Figure~\ref{fig:ResultHelium} shows a comparison to model predictions and to previous measurements \cite{Jaros78,Ableev85,Tanihata85,AMSXS}. Limited cross section measurements exist for helium-4 ions on heavy materials above 1~GeV. Our results are therefore compared to a light carbon target, for which more measurements are available. As in the case of proton, data points for carbon targets are rescaled to a Bi$_4$Ge$_3$O$_{12}$ target based on the cross section ratio in the EPOS LHC model. Figure~\ref{fig:ResultHelium} shows that good agreement is observed between our and previous measurements. The increase in cross section as a function of energy is observed to be slightly stronger than the model predictions, though within the uncertainties of the analysis. We find that the GLISSANDO3 model undershoots the measurement over the full energy range. At the lowest energy, EPOS LHC and QGSJetII-04 agree slightly better with the measurement than DPMJET-III due to their lower normalisation. Combining observations of proton and helium-4, EPOS LHC thus offers the best overall agreements to our measurements.

\begin{figure}[t]
	\includegraphics[width=8.6cm]{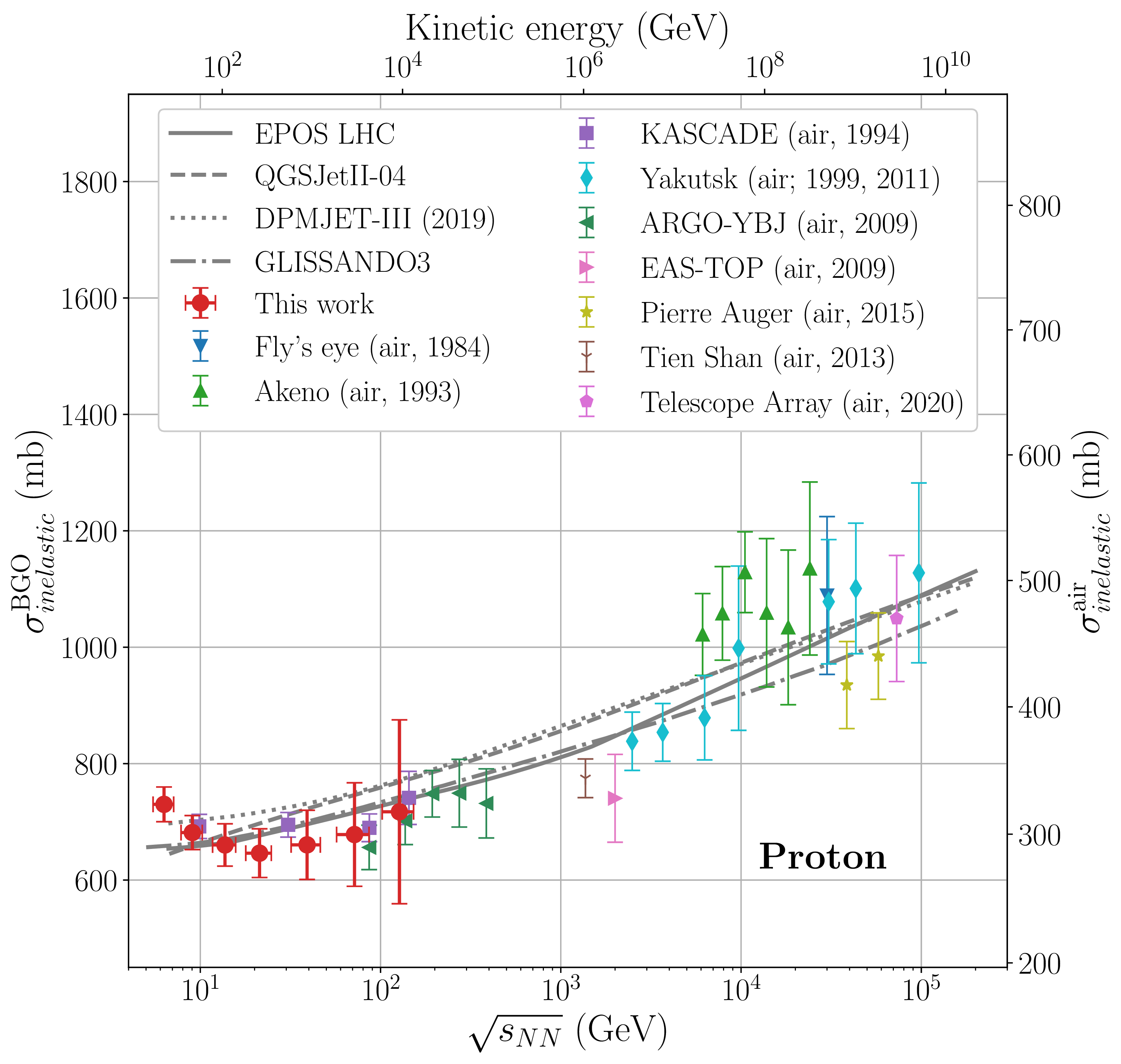}
	\caption{\label{fig:EASComparison} Comparison between our results for proton on BGO, the measurements from EAS experiments for proton-air interactions \cite{FlysEye,EASTOP,1999Yakutsk,2011Yakutsk,Akeno,TienShan,Auger,Auger2015,TelescopeArray,ARGO,KASKADE}, and model predictions \cite{EPOS,EPOS-LHC,dpmjet3_1,dpmjet3_2,dpmjet3_3,Glissando3}.}
\end{figure}

\textbf{EAS measurements.} Extensive Air Shower (EAS) experiments have probed the inelastic proton-air cross section over a kinetic energy range that spans more than eight orders of magnitude \cite{FlysEye,EASTOP,1999Yakutsk,2011Yakutsk,Akeno,TienShan,Auger,Auger2015,TelescopeArray,ARGO,KASKADE}. Air mainly consists of Nitrogen ($A=14$), making it a target material that is very different from Bi$_4$Ge$_3$O$_{12}$ for which proton interactions are mostly with bismuth ($A=209$). A direct comparisons between proton-BGO and proton-air measurements, as shown in Fig.~\ref{fig:EASComparison}, is thus more dependent on the assumed scaling model than in the case of lead ($A=207$). DPMJET-III was chosen to relate the BGO and air cross sections, as its scaling factor was in between that of QGSJetII-04, EPOS LHC and GLISSANDO3, with deviations up to 6.5\%. Measurements by ARGO-YBJ \cite{ARGO} and KASCADE \cite{KASKADE} in the same energy range are in excellent agreement with our results. In accordance they indicate that model predictions have a tendency to overestimate the inelastic cross section in the energy range $30\text{ GeV}\lesssim \sqrt{s_{NN}}\lesssim 3 \text{ TeV}$.

\textbf{CR flux normalisation.} Accurate cross section measurements are of vital importance for the precise measurement of CR fluxes. Given the observed count $N$ of incident CR particles, the flux averaged over an energy bin of width $\Delta E$ is given by
\begin{equation}
	\Phi(E\rightarrow E+\Delta E) = \frac{N}{\mathcal{A}_{eff}\cdot \Delta E \cdot \delta t},
	\label{eq:Flux}
\end{equation}
where $\mathcal{A}_{eff}$ is the effective detector acceptance in units m$^2$~sr and $\Delta t$ the live time of the measurement. For calorimetric experiments such as DAMPE, the energy of incident CRs can only be measured if the CR at some point interacts inelastically in the detector. Standard selection criteria therefore ensure that an inelastic interaction occurs, e.g. by imposing a lower limit on the total deposited energy. The result of applying these criteria is that the effective acceptance scales, to first order linearly, with the inelastic cross section. It follows from Eq.~\eqref{eq:Flux} that the overall normalisation of the flux will be too low (high), if the modelled cross section is too high (low). Since the calorimeter forms the bulk of detector material, this effect is strongly dependent on accurately modelling the cross section of the calorimetric material, in our case Bi$_4$Ge$_3$O$_{12}$.

\begin{figure}[t]
	\includegraphics[width=8.6cm]{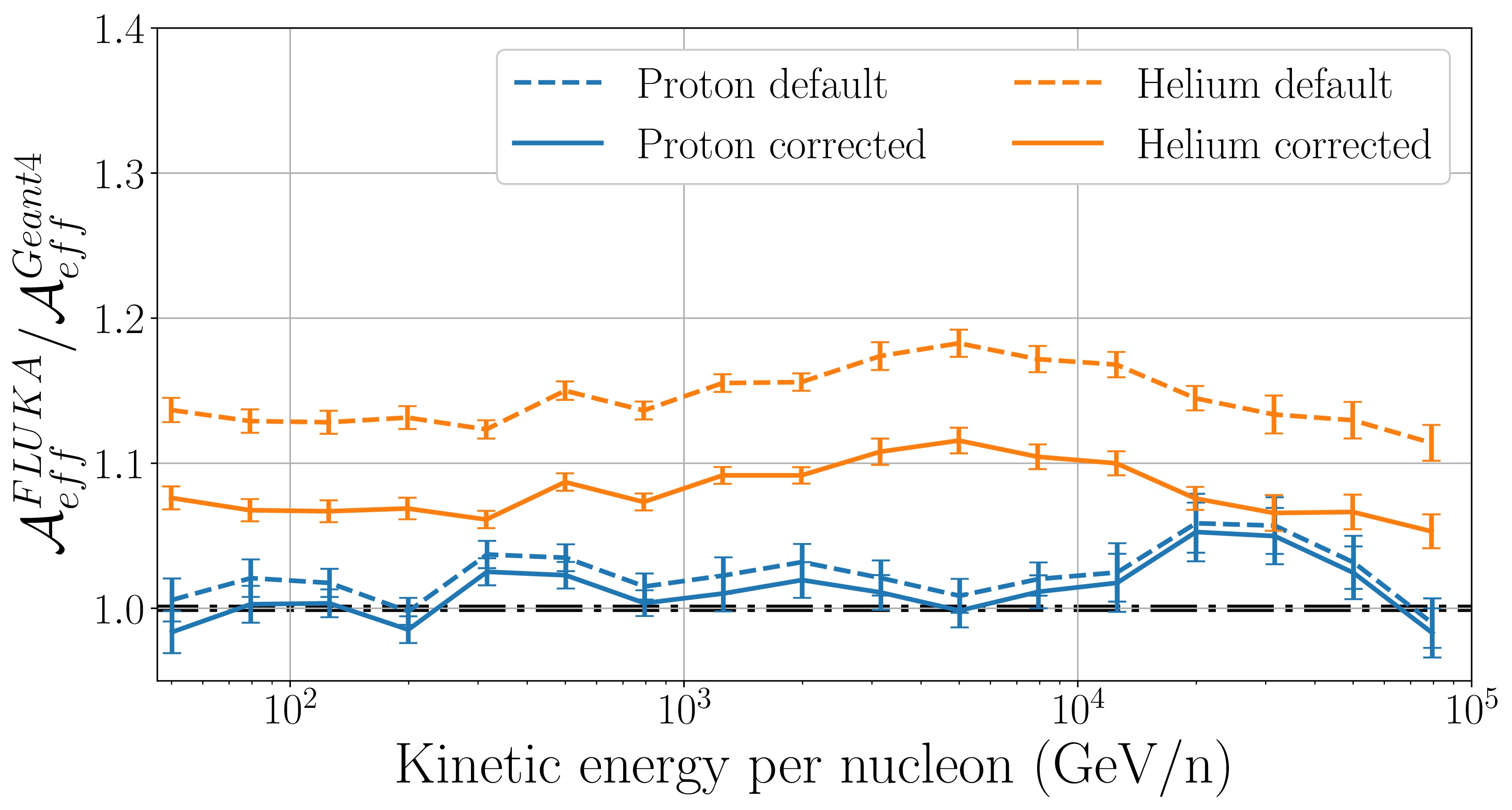}
	\caption{\label{fig:Acceptance} Ratio of the effective acceptance in DAMPE analyses of the proton and helium flux, when comparing FLUKA to Geant4. Dashed and full lines show the default ratio and cross section corrected ratio, respectively. When including the correction from the energy integrated analysis, a better overall agreement is observed between the two simulation frameworks.}
\end{figure}

Figure~\ref{fig:Acceptance} shows the ratio of $\mathcal{A}_{eff}$ between Geant4 and FLUKA for an analysis of the proton and helium flux with DAMPE. Dashed lines indicate the default ratio, while full lines show the ratio when the cross section in MC has been scaled to the measured result. The default ratio for proton is within 5\% at the considered energies, with an average deviation of 2.3\%. After the correction the ratio is systematically lower and closer to unity, with an average deviation of 1.2\%. A much stronger effect is observed for helium-4. FLUKA by default has a cross section which is 12\% to 14\% larger than that of Geant4 for kinetic energies between 10~GeV/n and $10^5$~GeV/n. The default acceptance from FLUKA is on average 14.5\% larger than that of Geant4, in large part due to its higher cross section. After setting the cross sections of both frameworks to the measured result, the average difference in acceptance decreases to 8.1\%, exhibiting a significant improvement. Roughly half of the discrepancy of the Geant4-FLUKA flux normalisation can thus be resolved by correcting the BGO cross section. Since the cross section of other detectors components were not corrected, we do not yet expect a perfect agreement. Additionally, differences in the physics model, particularly the multiplicity and energy distribution of secondary particles, are expected to cause the effective acceptance of Geant4 and FLUKA to deviate from each other even after correcting all cross sections. Combined, these effects are expected to account for the remaining 8.1\% differences.

Comparisons between the helium flux measurements of different experiments (see e.g. \cite{ArshiaICRC}) show that the spectra generally follow the same energy dependence, but are shifted by a constant normalisation factor. These shifts are within the uncertainty of the hadronic models, which forms the dominant systematic. Measurements from different experiments ~\cite{AMS7years,ATICHelium,CALETHelium,CreamHelium,ArshiaICRC,DAMPEProton,DAMPEHelium} typically rely on different simulation models. These include Geant4 (AMS-02, DAMPE), EPICS (CALET) \cite{EPICS1,EPICS2}, and FLUKA (ATIC, CALET, CREAM, DAMPE) . Figure~\ref{fig:Acceptance} suggests that the observed discrepancies in the normalisation of the helium flux can be largely due to the different hadronic simulation models which are used to interpret the measurements. Improving the precision of CR flux measurements and resolving these discrepancies necessitates improving the accuracy of hadronic models. A better knowledge of inelastic cross sections, as presented in this work, forms an important step towards reaching this goal.

\section{Conclusions \& Outlook}
In this work we presented the first cross section measurements performed with the DAMPE experiment. Eighty-eight months of data were analysed to study the interaction of proton and helium-4 CR ions on a Bi$_4$Ge$_3$O$_{12}$ target. An unbinned likelihood analysis was used to determine the value of the inelastic hadronic cross section. In the case of proton, the cross section was measured from 19~GeV to 9~TeV, while for helium the measurement spanned an energy range from 5~GeV/n to 3~TeV/n. These are the first cross section measurements for proton on Bi$_4$Ge$_3$O$_{12}$, and the first measurement for helium-4 ions at these energies for any heavy target material. Accurate measurements of the proton and helium-4 cross section form an important step towards improving the accuracy of CR flux measurements. We demonstrated that applying a correction for the measured cross sections improves the discrepancies in CR flux normalisation of proton and helium by 1.1\% and 6.4\%, respectively.

The methods developed for this analysis can be extended to other CR ions. Particularly CR particles such as carbon and oxygen are good targets due to their relatively high fluxes. Such analyses would require a significant modification of the event selection criteria due to the saturation effect of high-charge particles in the DAMPE tracker. We foresee to extend the cross section measurements to more ions in the future. On longer timescales, we note the potential of future space missions to extend the cross section measurement of CR ions to higher energies. In particular, the planned HERD experiment \cite{HERD} will increase the rate at which high-energy ions are detected by more than 1 order of magnitude, enabling cross section measurements at energies which are unreachable by the current generation of space experiments.

\begin{acknowledgments}
The DAMPE mission was funded by the strategic priority science and technology projects in space science of the Chinese Academy of Sciences (CAS). In China, the data analysis was supported by the National Key Research and Development Program of China (No. 2022YFF0503302) and the National Natural Science Foundation of China (Nos. 12220101003, 11921003, 11903084, 12003076 and 12022503), the CAS Project for Young Scientists in Basic Research (No. YSBR061), the Youth Innovation Promotion Association of CAS, the Young Elite Scientists Sponsorship Program by CAST (No. YESS20220197), and the Program for Innovative Talents and Entrepreneur in Jiangsu. In Europe, the activities and data analysis are supported by the Swiss National Science Foundation (SNSF), Switzerland, the National Institute for Nuclear Physics (INFN), Italy, and the European Research Council (ERC) under the European Union’s Horizon 2020 research and innovation programme (No. 851103). The authors gratefully acknowledge Dr. Alberto Ribon for engaging in fruitful discussions and for providing valuable feedback during the course of this work.

\end{acknowledgments}

\vfill\null

\appendix

\section{Lepton rejection\label{app:LeptonRejection}}

The DAMPE experiment is continually bombarded by a variety cosmic-ray (CR) particles. For the analysis of the proton and helium-4 inelastic cross section, CR leptons form a small yet non-negligible background. Events induced by CR photons can easily be removed due to their lack of charge deposited in the plastic scintillator detector (PSD). Events induced by CR electrons and positrons, however, are more difficult to reject as they deposit a charge signal identical to that of proton. To distinguish protons from charge $\pm e$ leptons, a variable has been developed in previous work \cite{DampeElectronNature} which probes the collimation and depth of the particle shower in the bismuth-germanium-oxide (BGO) calorimeter. 

\textbf{Collimation.} Each of the 14 layers of the BGO calorimeter comprises 22 adjacent bars. The root-mean-square (RMS) spread of energy within layer $j$ is defined as
\begin{equation}
	\text{RMS}_j = \sqrt{ \frac{\sum_{i=1}^{22}E_{ij}\cdot \left(x_{ij}-x_{j}^{COG}\right)^2}{\sum_{i=1}^{22}E_{ij}} },
	\label{eq:RMS}
\end{equation}
where $E_{ij}$ is the energy measured in bar $i$ of layer $j$, $x_{ij}$ is the horizontal coordinate of that bar, and $x_{i}^{COG}$ is the center-of-gravity (COG) in layer $j$. The COG is defined as:
\begin{equation}
	x_{j}^{COG} = \begin{cases}
		x_{mj} & \text{if $m=1$ or $m=22$}, \\
		\frac{\sum_{i=m-1}^{m+1}E_{ij}\cdot x_{ij}}{\sum_{i=m-1}^{m+1}E_{ij}} & \text{if $2\leq m \leq 21$},\\
	\end{cases}       
\end{equation}
where $m$ is the index of the bar with the maximal energy measured in layer $j$.

\textbf{Depth.} As a measure of how far the particle shower penetrated in the BGO calorimeter, we consider the ratio
\begin{equation}
	F_L = E_L / E_{tot},
\end{equation}
where $E_L$ is the energy measured in the last layer of the calorimeter which has non-zero energy, and $E_{tot}$ is the total energy summed over all 308 bars of the calorimeter.

\begin{figure}[h]
	\includegraphics[width=8.6cm]{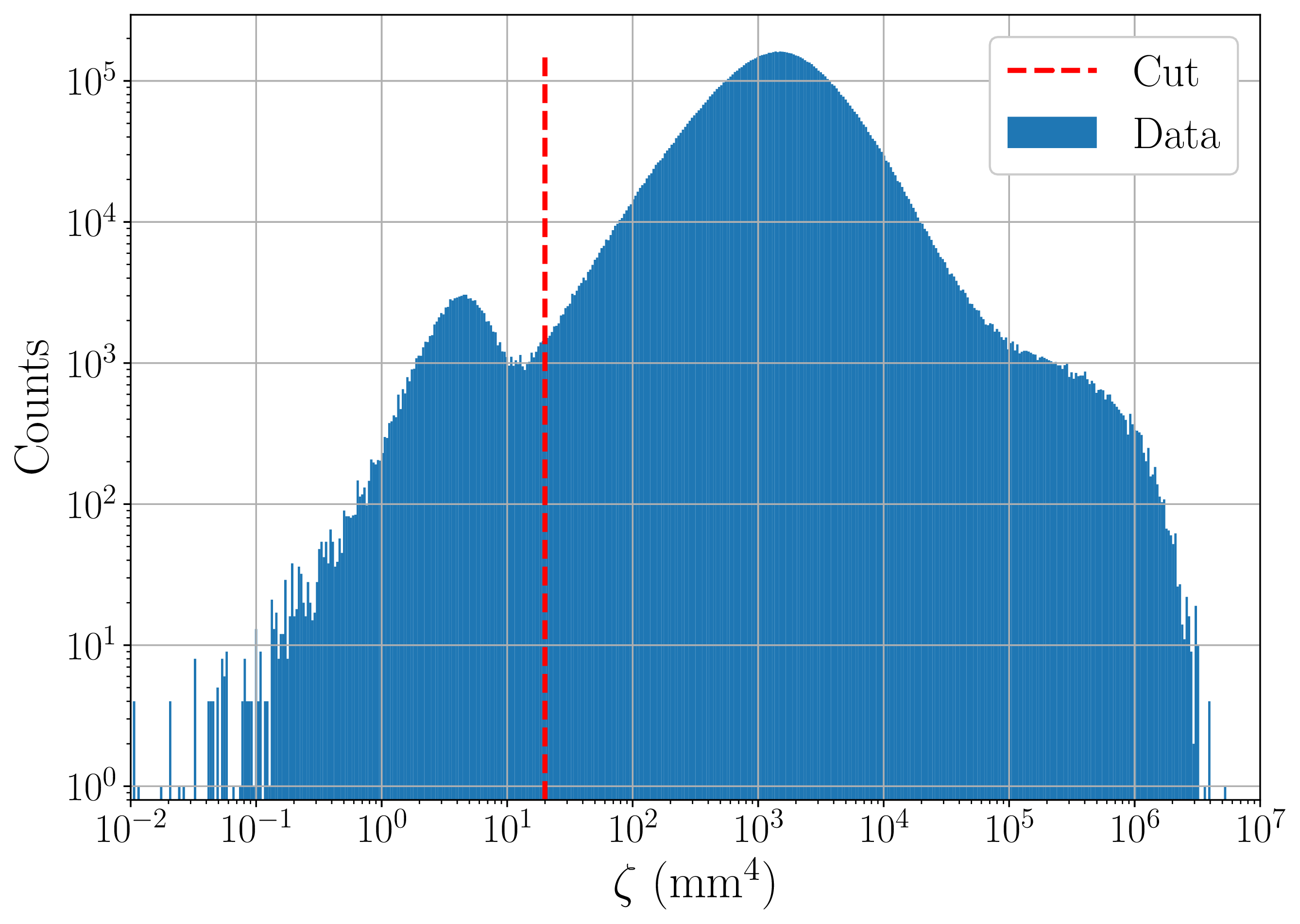}
	\caption{\label{fig:XTRL} Distribution of the $\zeta$ variable, defined in Eq.~\eqref{eq:XTRL}, for data after events applying the containment cuts. The small peak on the left is due to the leptonic background, mostly comprised of electrons. A conservative cut $\zeta>20\text{ mm}^4$ is imposed to remove these events from our analysis.}
\end{figure}

\textbf{Combined.} Electromagnetic showers are more collimated, and shorter in comparison to hadronic showers. For these reasons, both RMS$_j$ and $F_L$ are smaller for electromagnetic showers. As presented in previous work \cite{DampeElectronNature}, the product
\begin{equation}
	\zeta = \left( \sum_{j=1}^{14} \text{RMS}_j\right)^4\cdot F_L,
	\label{eq:XTRL}
\end{equation}
shows a strong separation power to distinguish electromagnetic from hadronic showers. The distribution of $\zeta$ in data after containment cuts is shown in Figure~\ref{fig:XTRL}. Two contributions are clearly distinguishable from leptons (left) and hadrons (right). A cut $\zeta>20\text{ mm}^4$ is used to remove the leptonic background. 

\section{Interaction depth classifier\label{app:Vertex}}

The cross section measurements in this work rely on predicting the depth at which particles interact inelastically in the calorimeter. This prediction is made using an XGBoost (XGB) classifier \cite{XGB}. A brief description of the input and output parameters of the classifier is provided in the main article. In this section, additional details are provided on the classifier training and performance.

\textbf{Training.} Simulations were carried out for proton and helium-4 with both Geant4 and FLUKA. Prior to training, the same event selection described in the paper was applied to the Monte Carlo (MC) events, with two distinctions.
\begin{enumerate}\addtolength\itemsep{-2.5mm}
	\item The true direction of the primary was used to determine if the event is fiducially contained, rather than the reconstructed STK track.
	\item No cut was placed on the median STK charge, to enable selecting both proton and helium at the same time.
\end{enumerate}
Table~\ref{tab:MCsamples} shows the energy range for each of the four samples, and their statistics after the event selection. Roughly equal number of simulated events are used in the training process, except for proton-FLUKA which was limited by the total size of the simulation set.

\begin{table} %Star makes the table span the full page width
	\caption{\label{tab:MCsamples} Energy range of the four samples used to train the XGB classifier. Counts correspond to the number of events that were used in the training process.}
	\begin{ruledtabular}
		\begin{tabular}{cccc}
			Sample&Framework& Energy range &\text{Counts}\\
			\hline
			Proton&Geant4&10 GeV - 1 PeV& 3,171,334 (35.8\%) \\
			Proton&FLUKA&10 GeV - 1 PeV& 517,121 (5.8\%) \\
			Helium&Geant4& 10 GeV - 500 TeV & 2,757,538 (31.1\%) \\
			Helium&FLUKA& 10 GeV - 500 TeV & 2,417,593 (27.3\%) \\
		\end{tabular}
	\end{ruledtabular}
\end{table}

\begin{figure}[h]
	\includegraphics[width=8.6cm]{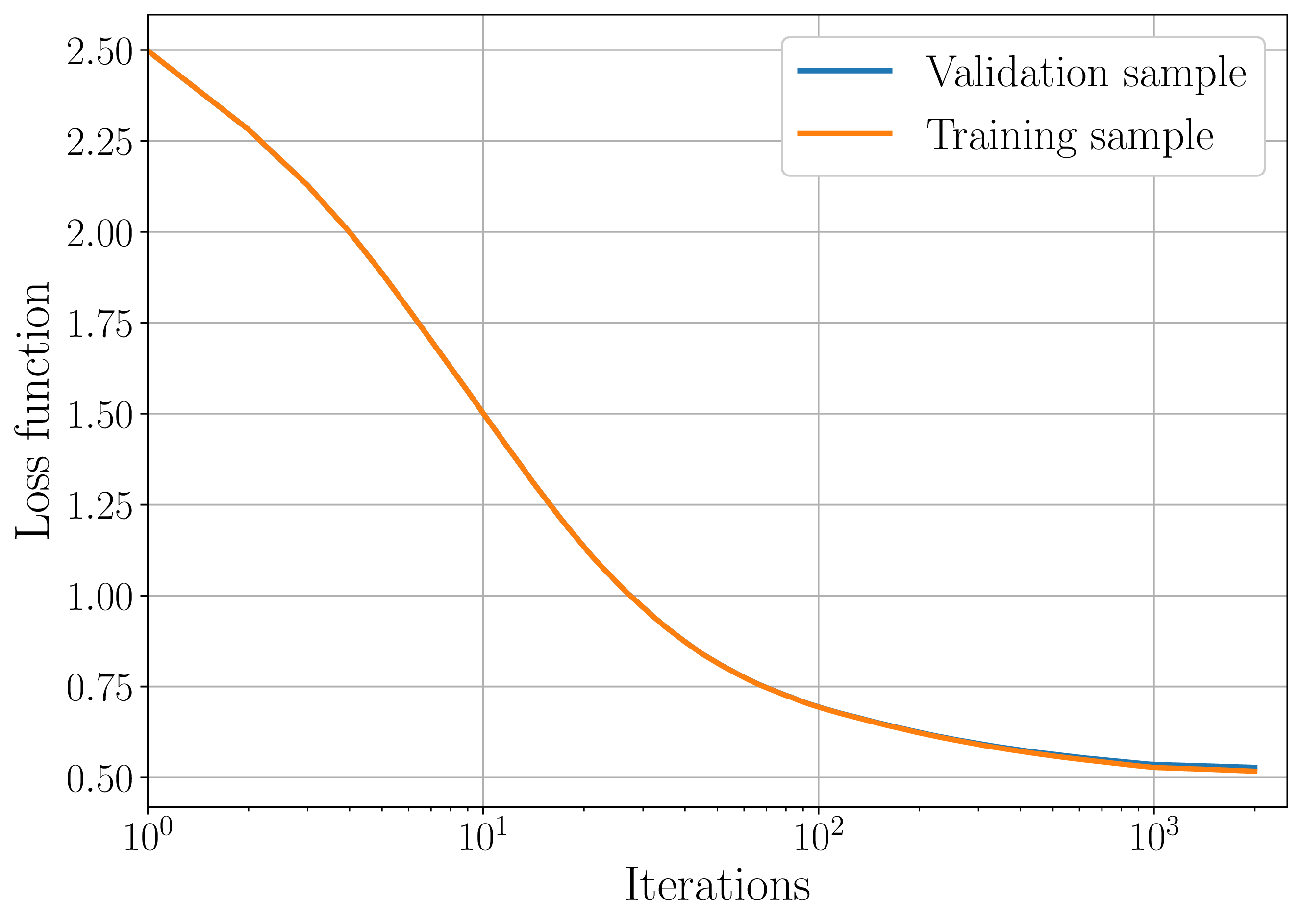}
	\caption{\label{fig:ClassifierTraining} Value of the multi-class loss function as a function of the number of training steps. The training sample is shown to have converged, while also remaining consistent with the validation sample.}
\end{figure}

Good convergence was observed when setting the XGB hyperparameters to: a maximal tree depth of 0.8, a subsampling of 0.8, a column subsampling of 0.5 per tree, and the softmax training objective. Figure~\ref{fig:ClassifierTraining} shows the loss function as a function of the number of trees or iterations. A learning rate of 0.1 was used for the first half of the training, which was lowered to 0.03 for the latter half. After two thousand iterations, the model is observed to stagnate, with minimal difference between the loss function of the training and validation sample.

\begin{figure}[h]
	\includegraphics[width=8.6cm]{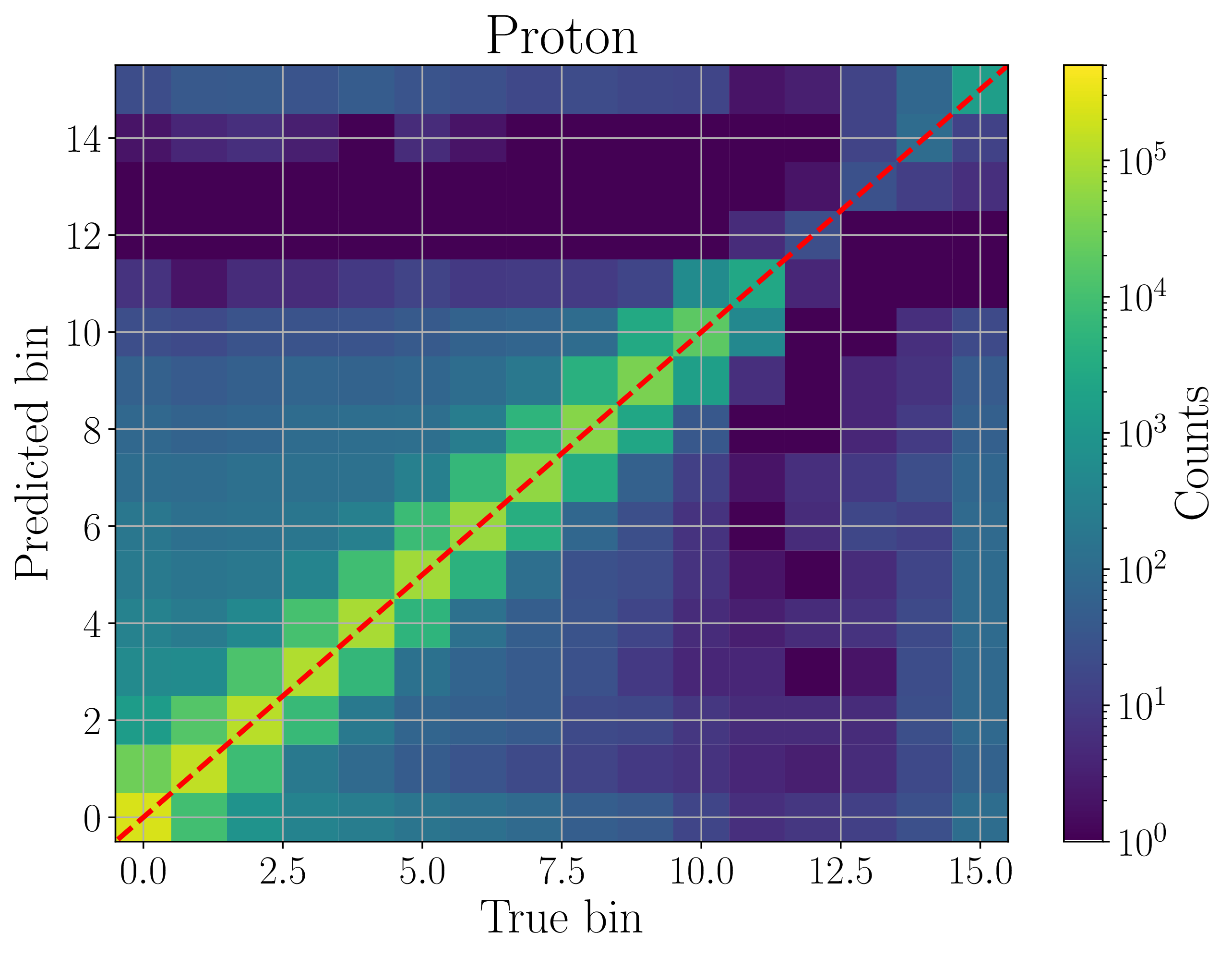}
	\caption{\label{fig:Classifier2d} Correlation between the true category to which an event belongs, and the category predicted by the classifier. The colour scale shows the number events, in this case proton simulated with Geant4.}
\end{figure}

\begin{figure}[h]
	\includegraphics[width=8.6cm]{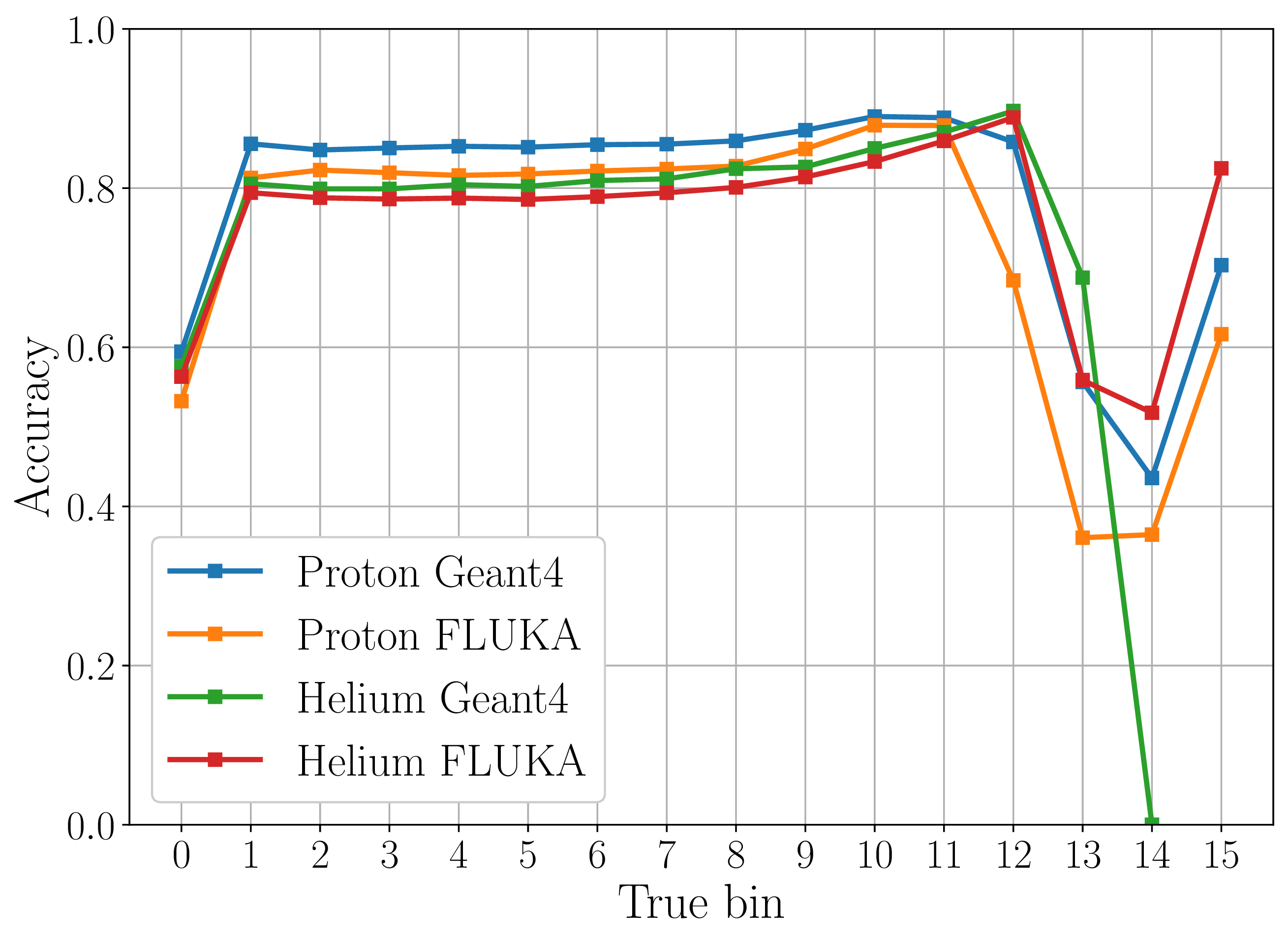}
	\caption{\label{fig:Classifier1d} Fraction of events for which the classifier correctly predicts the category, as a function of the true category. Excluding the first two bins and grouping the last 6 bins, an accuracy of $\geq$80\% is obtained in all cases.}
\end{figure}

\textbf{Performance.} Figure~\ref{fig:Classifier2d} shows the accuracy of the trained classifier on proton simulated with Geant4 (similar results are obtained for helium and FLUKA). Good convergence is observed. Misclassified events also shown to be most likely classified into an adjacent bin. The fraction of events that is correctly classified per truth bin is shown in Figure~\ref{fig:Classifier1d} for all four categories. As noted in the main text, it can be challenging to differentiate between events that interact before or in the first layer of BGO. A lower accuracy is therefore observed in the first bin. Following the first two bins, a consistent accuracy is observed above 78\% for all four samples. After the 10th bin, a drop in accuracy is again observed due to the very limited number of events that interact in the last layers of BGO. For this reason, the last six categories are grouped into a single bin in the analysis.

\begin{figure}[h]
	\includegraphics[width=8.6cm]{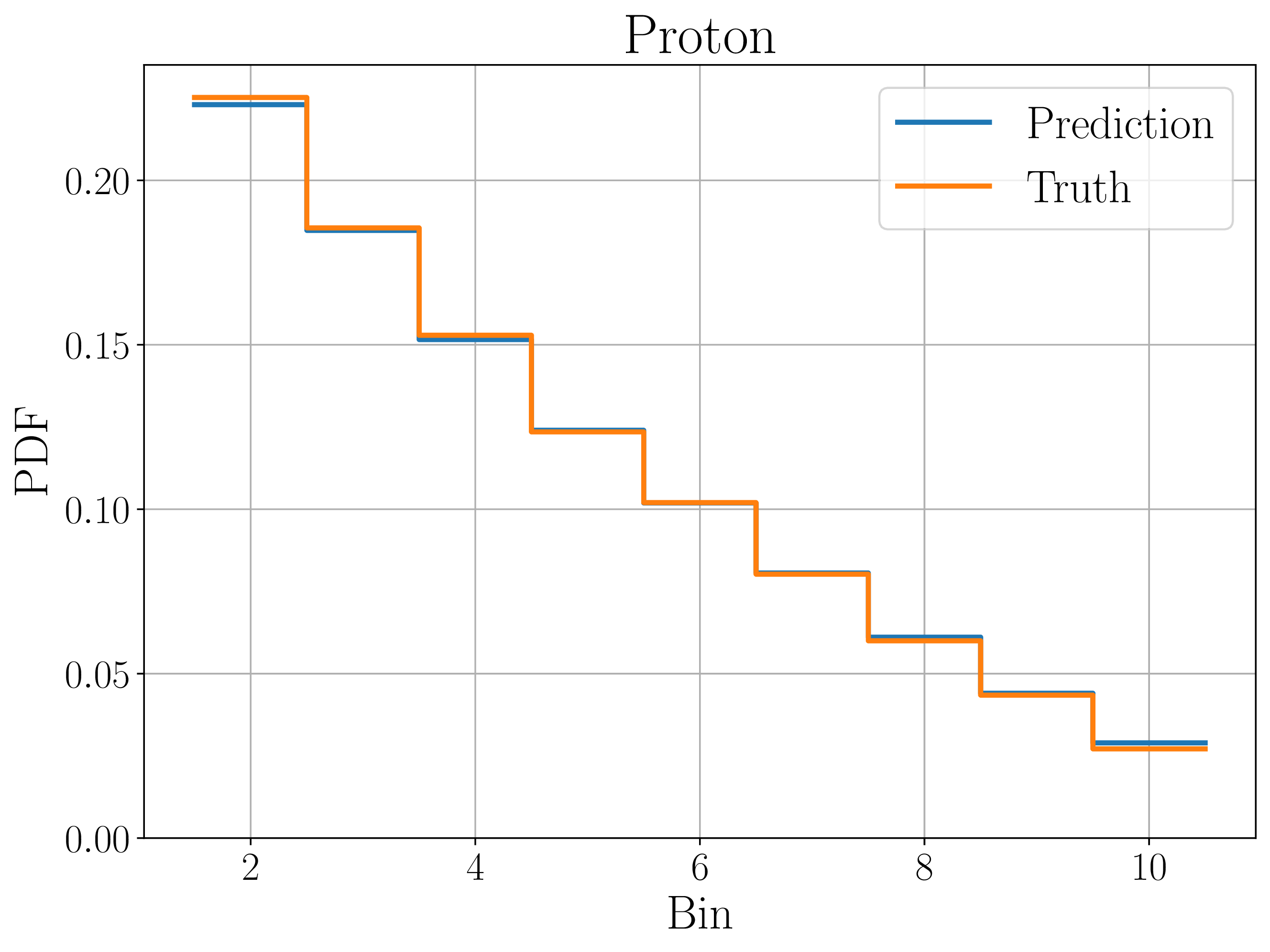}
	\caption{\label{fig:ClassifierBias} Fraction of true and predicted events in each category for proton simulated with Geant4. An excellent agreement is observed, indicating that there is a minimal bias between the true and predicted number of events per class.}
\end{figure}

Aside from a high accuracy, an important requirement of the classifier is that it does not introduce any bias. If misclassified events are more likely to fall before/after the true bin, this would shift the observed distribution, potentially leading to artificially higher or lower cross sections. Figure~\ref{fig:ClassifierBias} shows the ratio $N_i/\sum_{j=2}^{10}$ for proton simulated with Geant4. Little to no bias is observed. Similar results hold for helium and simulation with FLUKA. To mitigate the potential impact of any minimal yet non-zero bias on our measurements, the likelihood procedure also uses the predicted rather than true number of events per category for MC, to allow a fair comparison to data for which only the prediction is available.

\section{Cross section re-weighting\label{app:Reweight}}

\textbf{Method.}
A procedure has been developed to re-weight existing samples to different cross sections. The computationally laborious process of re-simulating MC samples over a grid of cross sections can thus be avoided. Modifications to the cross section effectively change the depth at which particles interact inelastically in the detector. A scale factor, $\kappa$, is introduced to quantify the change in cross section
\begin{equation}
	\sigma_{true}(E) = (1+\kappa )\cdot \sigma_{MC}(E).
	\label{eq:Scaling}
\end{equation}
Prior to any modification $(\kappa=0)$, the probability density for a particle to interact at a depth $z$ is given by
\begin{equation}
	\phi(z,\kappa=0) = \frac{d\mathcal{P}}{dz}(z,\kappa=0).
\end{equation}
The cumulative fraction of particles that interact inelastically before reaching a depth $z$ is given by
\begin{equation}
	\Phi(z,\kappa=0) = \int_{z_{min}}^{z}dz' \phi(z',\kappa=0),
\end{equation}
where $z_{min}$ marks the start of the BGO calorimeter. Assuming a simplified scenario in which all particle are vertically down-going and have the same energy, this cumulative distribution will follow an exponential decay:
\begin{equation}
	\Phi(z,\kappa=0) = 1-\exp\left(n\cdot \sigma\cdot z\right),
	\label{eq:ExpDecay}
\end{equation}
where $n$ is the total number of bismuth, germanium, and oxygen nuclei per unit volume; and $\sigma$ is the cross section. It follows from Eq.~\eqref{eq:ExpDecay} that a change to the cross section $(\kappa \neq 0)$ leads to the following change in the cumulative distribution:
\begin{equation}
	\Phi(z,\kappa) = 1-\left[1-    \Phi(z,\kappa'=0)     \right]^{1+\kappa}.
\end{equation}
From this result, $\Phi(z,\kappa)$ can be derived with respect to $z$ to find the value of $\phi(z,\kappa)$. Assuming an MC sample which has been simulated with $\kappa'=0$, it follows that the weighting factor required to modify the cross section by a fraction $\kappa$ is given by
\begin{equation}
	w(z) = \frac{\phi(z,\kappa)}{\phi(z,\kappa'=0)}.
\end{equation}

\begin{figure}[h]
	\includegraphics[width=8.5cm]{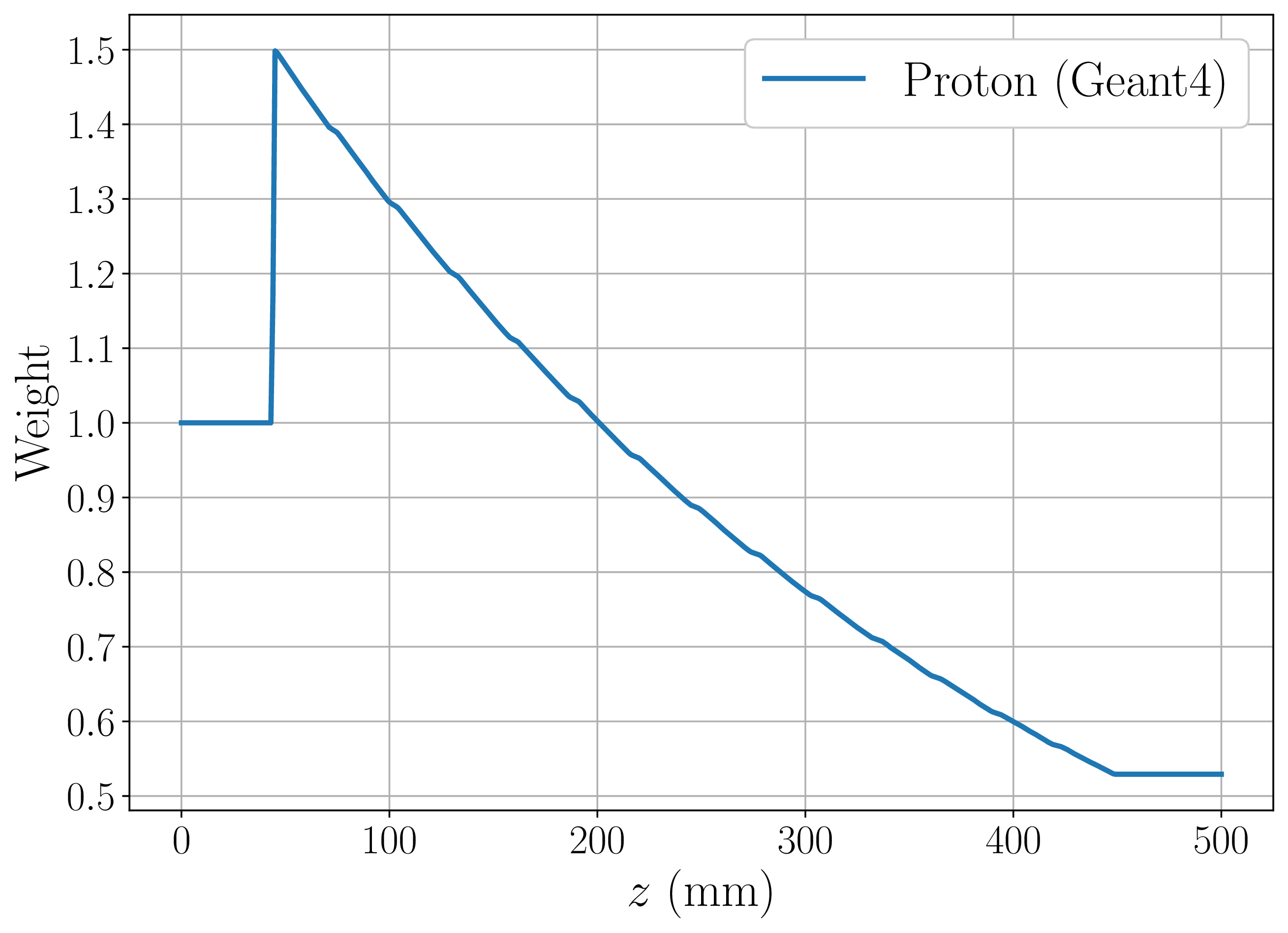}
	\caption{\label{fig:Weighting1d} Weights attributed to 1~TeV proton primaries in Geant4 when increasing the cross section by 50\%. Minor steps can be observed every 29~mm in the curve due to the gap in between the BGO layers.}
\end{figure}

Figure~\ref{fig:Weighting1d} shows an example of the weights simulated for proton interacting with BGO in Geant4 when $\kappa=0.5$. Prior to BGO ($z<44$~mm) the weights are unchanged. At the start of the calorimeter, a weight of 1.5 is obtained as particles are 50\% more likely to interact. Further along in the calorimeter, the weight drops as more initial interactions implies that there are less particles left to interact at a later stage. Below the calorimeter ($z>$~450~mm), the weight is constant again.

\textbf{Generalisation.}
The method described above applies only when all incident particles are of the same type, and when they have the same energy and incident angle. To generalise the procedure, events of a fixed primary type are binned as a function of energy and incident angle. For each category of events, numerical histograms are produced of the depth at which events interact inelastically. After applying the re-weighting procedure, a characterisation is obtained of $w(z)$ for that particular bin.

\begin{figure}[h]
	\includegraphics[width=8.6cm]{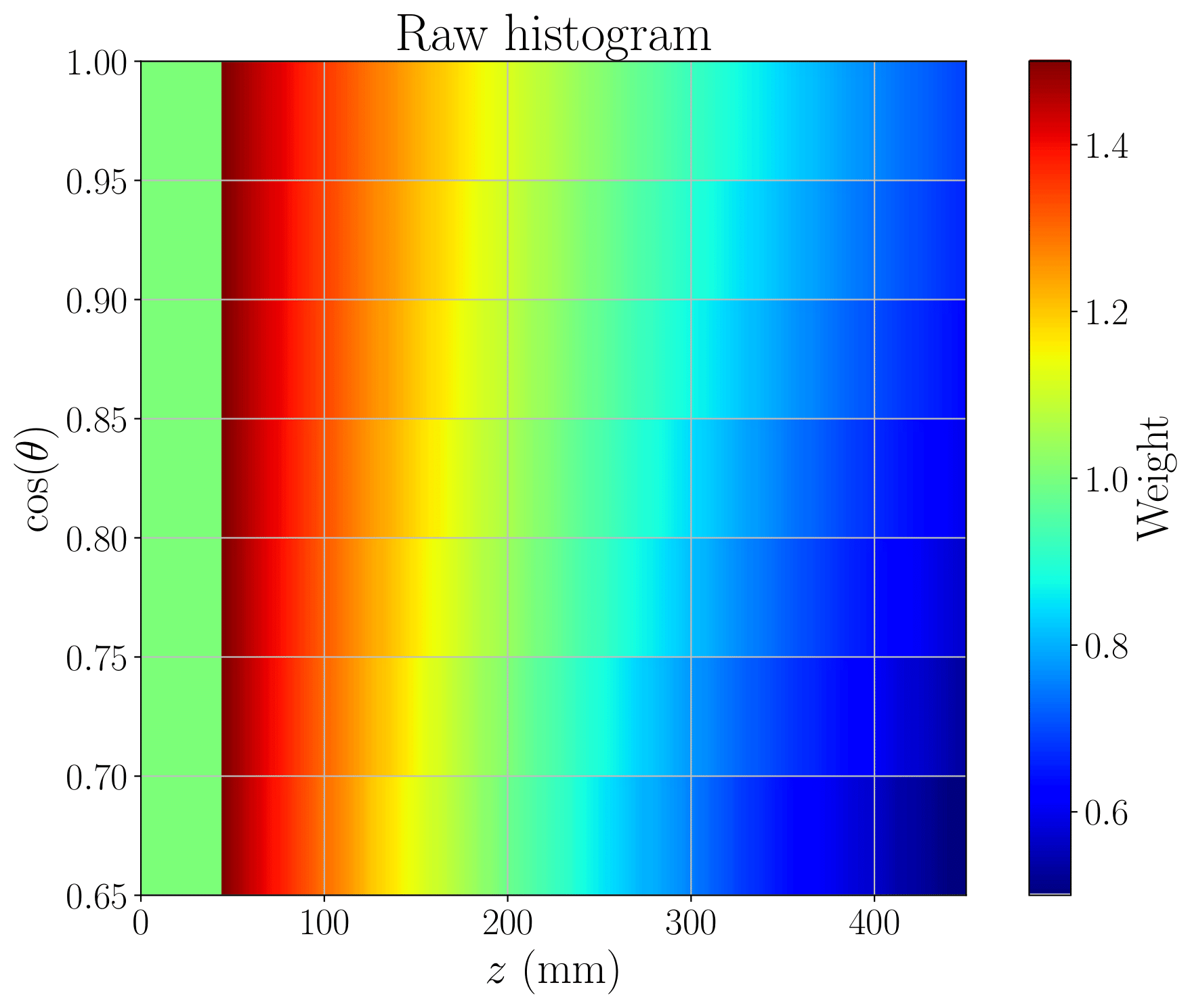}
	\includegraphics[width=8.6cm]{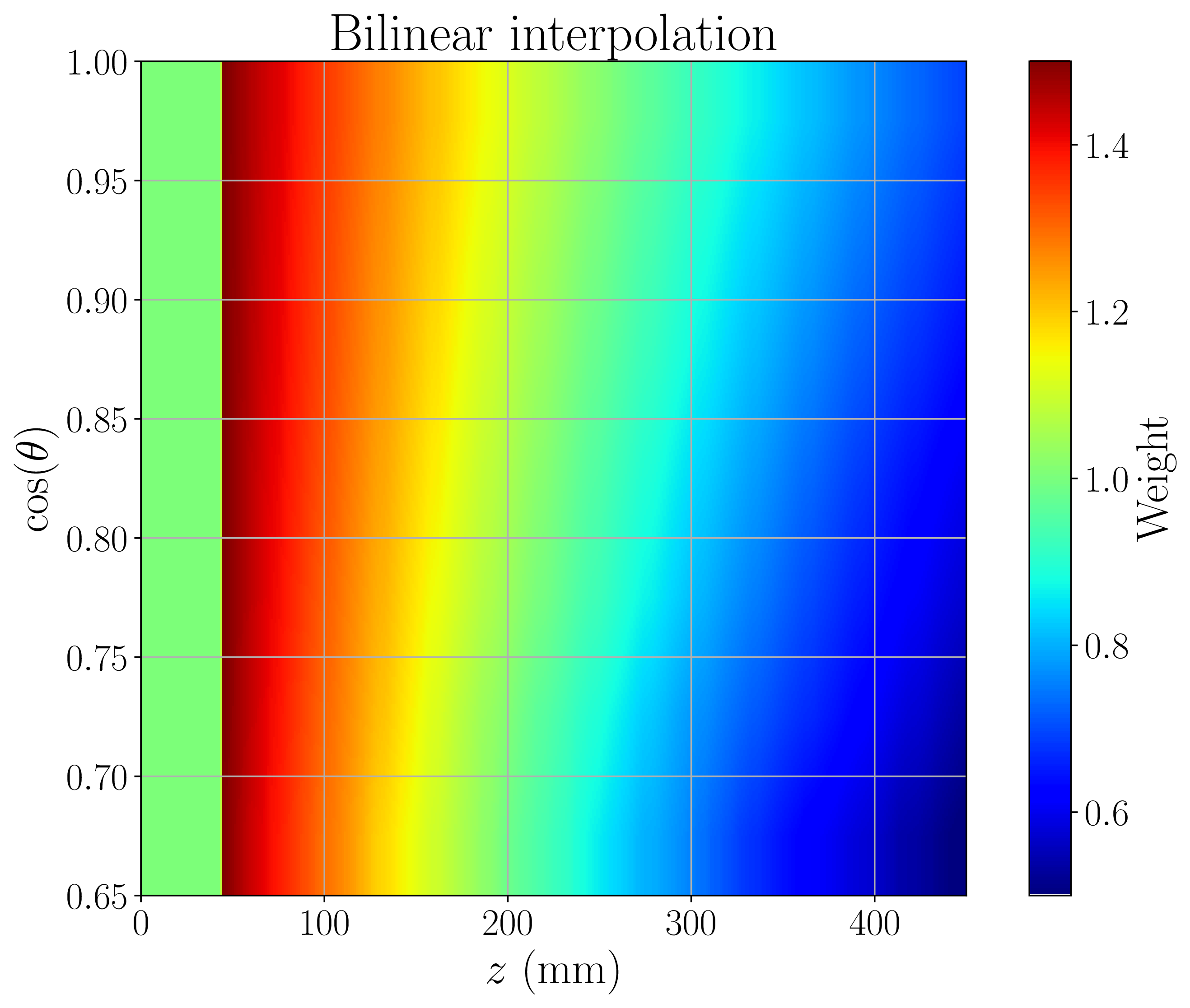}
	\caption{\label{fig:Weighting2d} Visualisation of how the raw 3 dimensional histograms $(E,\theta,z)$ are interpolated to enable computing the weight for arbitrary events. The top figure show the histograms for a single energy bin (10-31.6 GeV) in case of a proton primary when increasing the cross section by 50\%. The bottom figure shows the same histogram, but smoothened out using bilinear interpolation.}
\end{figure}

Weights of arbitrary events are computed by considering the true kinetic energy, incident angle, and interaction point of the event $(E,\theta,z)$; and making a 3D interpolating over the constructed histograms. Figure~\ref{fig:Weighting2d} visualises this procedure. For the sake of clarity the primary is assumed to be a proton particle with a fixed energy of 17.8~GeV. Hence, Figure~\ref{fig:Weighting2d} shows a 2D projection of the event weight as a function of $\theta$ and $z$.

\section{Energy dependent uncertainties\label{sup:Uncertainties}}

\begin{figure}[h]
	\includegraphics[width=8.6cm]{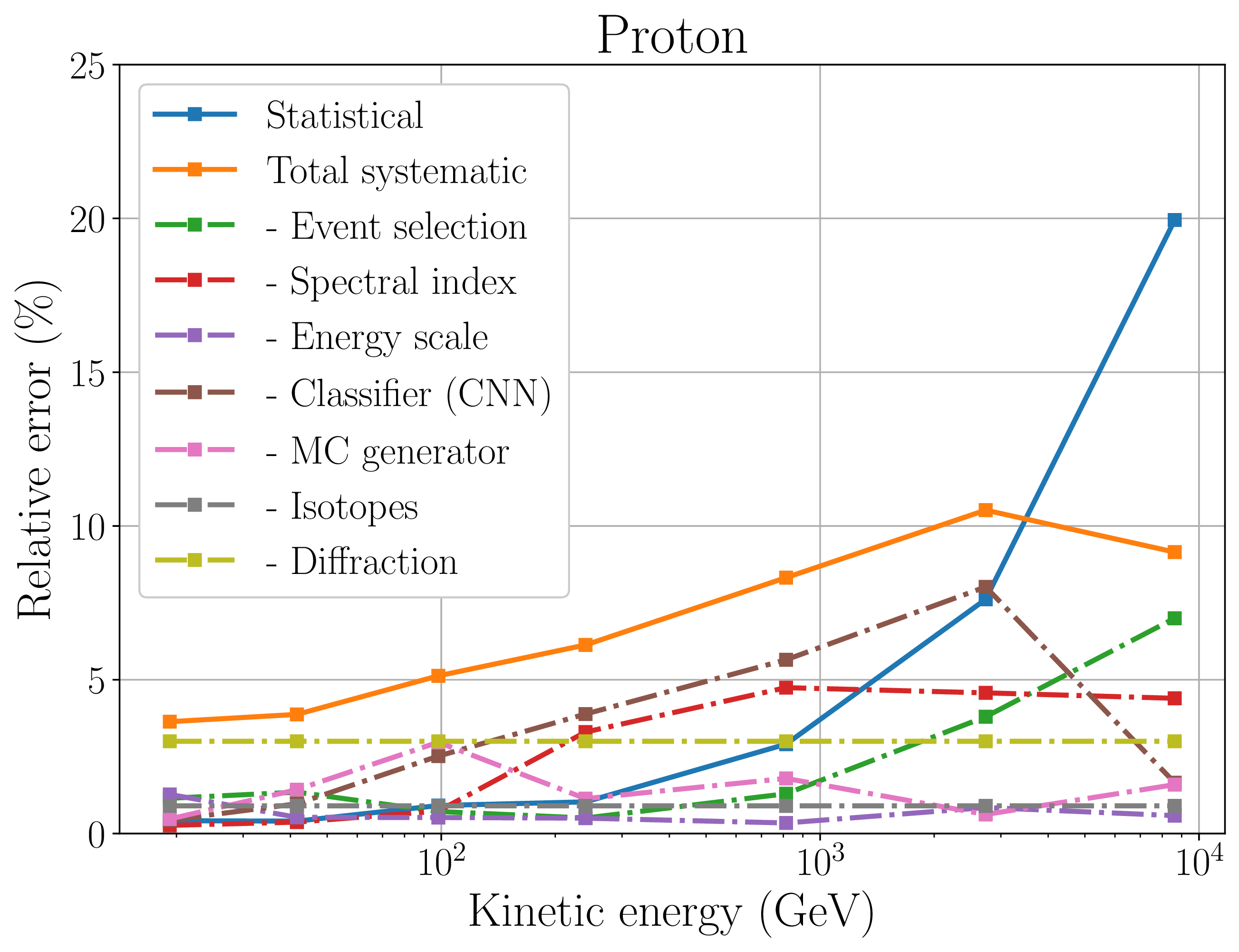}
	\caption{\label{fig:UncertaintiesProton} Overview of the relative systematic errors for each of the seven energy bins in the proton analysis.}
\end{figure}

\begin{figure}[h]
	\includegraphics[width=8.6cm]{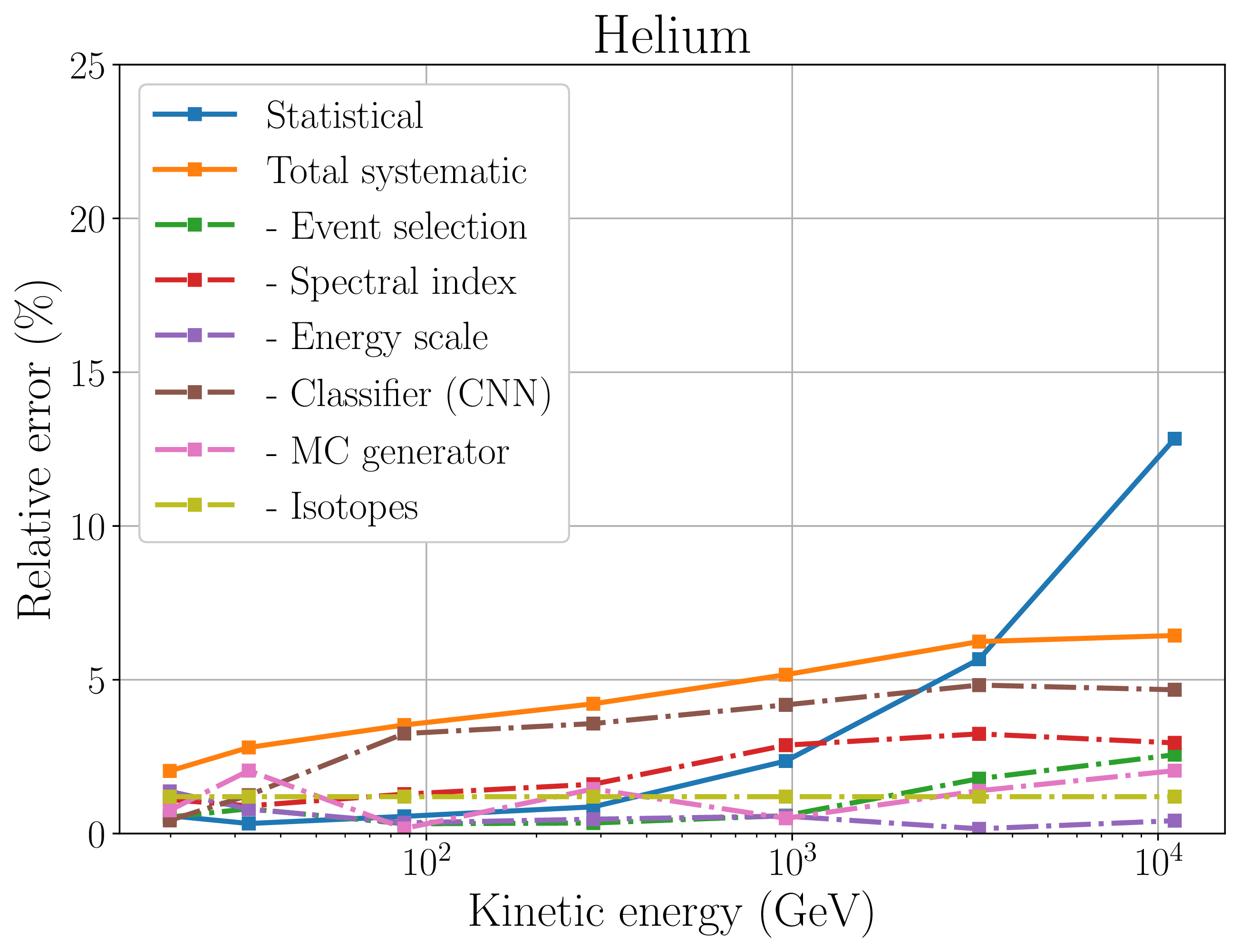}
	\caption{\label{fig:UncertaintiesHelium} Overview of the relative systematic errors for each of the seven energy bins in the helium-4 analysis.}
\end{figure}

Figure~\ref{fig:UncertaintiesProton} and \ref{fig:UncertaintiesHelium} show the relative statistical and systematic error of the analysis as a function of energy for proton and helium-4, respectively. The interaction depth classifier is found to be the dominant contribution to the total systematic error in 3 out of 7 bins for both proton and helium-4, showing a generally increasing trend as a function of energy. This increase is attributed in large part to energetic back-scattered particles in high-energy showers. Such particles can lead to significant energy depositions in the calorimeter above the shower, thus making it more challenging to determine the exact depth at which the inelastic interaction of the primary particle occurred.

An increasing trend as a function of energy is also observed for the systematic uncertainty related to the spectral index and event selection. This feature is by design in case of the spectral index. Very accurate observations of the CR flux at energies below the break at $\sim500$~GeV (proton) and $\sim$1~TeV (helium) enable smaller uncertainties on the spectral index used to weight MC than at higher energies. In case of the event selection, the effect is likely again related to the increased number of back-scattered particles in high-energy events. These backscatters make identifying the primary track in the STK detector more challenging, affecting both the selection efficiency and background of the candidate proton or helium-4 events selected from data.
%Helium is easier to recognise than proton (less background from backscattered particles in the tracker)

No distinctive energy dependence is observed when changing the MC generator, i.e. when comparing simulation produced by Geant4 and FLUKA. On the contrary, a dependence is seen for the energy scale, which has its largest uncertainty for the lowest energy bin. This effect is related to the shape of the input spectrum. At a rigidity of roughly 20~GV, a sharp break is observed in the spectrum due to the geomagnetic cut-off \cite{CLAY,CLAY2}. A relatively small mismodelling of the average energy that CRs deposit in the calorimeter could therefore have a non-negligible impact on our measurement in the lowest energy bin. At higher energies, spectral features are much less strong than the geomagnetic cut-off, and the uncertainty on the energy scale becomes $\ll$1\% .

\bibliography{Bibliography}% Produces the bibliography via BibTeX.

\end{document}